\documentclass[letterpaper, 10pt, journal]{ieeeconf}   

\IEEEoverridecommandlockouts                              
\usepackage{graphics} 
\usepackage{epsfig} 
\usepackage{mathptmx} 
\usepackage{times} 
\usepackage{amsmath} 
\usepackage{amssymb}  
\usepackage{subcaption}
\usepackage{float}
\usepackage[plain]{algorithm}
\usepackage[noend]{algpseudocode}
\usepackage{booktabs}
\makeatletter
\def\BState{\State\hskip-\ALG@thistlm}
\makeatother
\usepackage{fancyhdr}
\title{\huge \bf
A novel entropy-based hierarchical clustering framework for ultrafast protein structure search and alignment
}

\author{Baris Ekim$^{1,2}$
\thanks{$^{1}$Armand Hammer United World College of the American West (UWC-USA), Montezuma, NM 87731, USA}%
\thanks{$^{2}$Computer Science and Artificial Intelligence Laboratory (CSAIL), Massachusetts Institute of Technology (MIT),
        Cambridge, MA 02139, USA}%
}

\usepackage{pgfplots}
\usepackage{textcomp}

\begin{document}

\maketitle
\thispagestyle{plain}
\pagestyle{plain}
\pagenumbering{arabic}

\begin{abstract}

Identification and alignment of three-dimensional folding of proteins may yield useful information about relationships too remote to be detected by conventional methods, such as sequence comparison, and may potentially lead to prediction of patterns and motifs in mutual structural fragments. With the exponential increase of structural proteomics data, the methods that scale with the rate of increase of data lose efficiency. Hence, new methods that reduce the computational expense of this problem should be developed.
We present a novel framework through which we are able to find and align protein structure neighbors via hierarchical clustering and entropy-based query search, and present a web-based protein database search and alignment tool to demonstrate the applicability of our approach. The resulting method replicates the results of the current gold standard with a minimal loss in sensitivity in a significantly shorter amount of time, while ameliorating the existing web workspace of protein structure comparison with a customized and dynamic web-based environment. Our tool serves as both a functional industrial means of protein structure comparison and a valid demonstration of heuristics in proteomics.

\end{abstract}

\section{INTRODUCTION}

Predicting the function of a protein is key to understanding life at a molecular level. Conventionally, comparison of primary sequences of proteins has long been a widely-used method for detecting proteins that share a similar function and modeling phylogenetic trees [1, 2]. However, sequence comparison alone is not purposive for detecting distant evolutionary connections between proteins, which may elucidate useful functional information [3 - 5]. Sequence similarity, although simpler and more streamlined, is a far less accurate predictor of functional similarity than structure comparison, which is significantly more powerful for identifying cases where the evolutionary progress of the individual subject protein precludes the comparison of the protein sequences. Investigating structural similarity is also functional for analyzing cases with insufficient sequence similarity [6] or identifying mutual functional protein sites and motifs [7, 8]. 
\begin{figure}
\centering
\begin{minipage}{.22\textwidth}
\includegraphics[width=\linewidth]{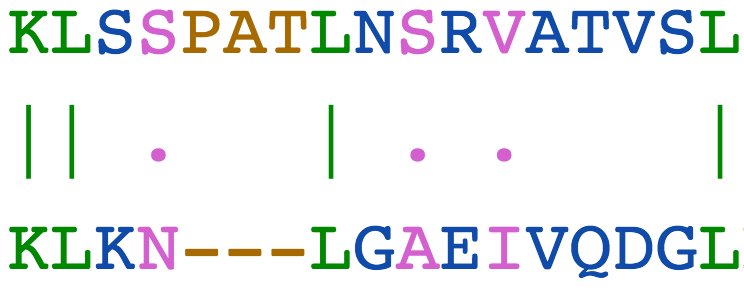}
\end{minipage}
\quad
\begin{minipage}{.22\textwidth}
\includegraphics[width=\linewidth]{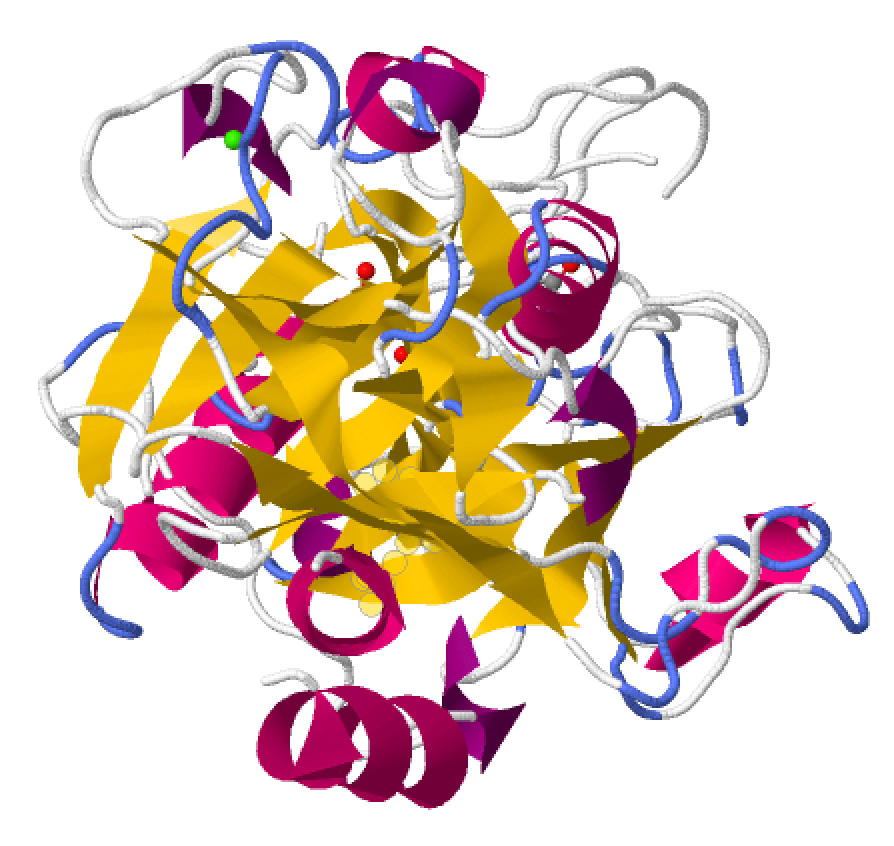}
\end{minipage}
\caption{Sequence (left) and structure (right) alignment of $\alpha$-trypsin (1AKS) and flavodoxin (1AKR).}
\end{figure}
Problems regarding similarity search also arise in many different applications in computer science, such as knowledge discovery and data mining (KDD) [9], vector quantization, and pattern recognition and classification [10]. Similarity search is not only fundamental to data science, but it also lies within problem domains in various other fields where data are not represented numerically, such as the statistical analysis of sets of molecular descriptors [11], investigation of cross-generational mutative patterns [12], and the representation of environmental sound waveform features [13]. \\\\
\indent In the recent years, an explosion of newly discovered protein structures has been witnessed. Consequently, the Protein Data Bank (PDB), the largest protein structure database available online, has come to surpass 110,000 structural entries, with an increase of around 7,000 in 2015. As the amount of information in the PDB becomes overwhelmingly high, the need for methods to organize and classify structures for structure neighbor identification arises [14]. 
\begin{figure}
\centering
\begin{tikzpicture}[trim axis left, trim axis right]
 \centering
\begin{axis}[
legend style={at={(0.5,-0.1)},anchor=north},
ylabel near ticks,
xlabel near ticks,
xmin=, 
xmax=, 
ymin=0,
ymax=1000000,
xtick=data,
xtick style={draw=none},
ylabel={Number of structures},
xlabel={Year},
xticklabels={},
]
\addplot table [scatter, x=Year, y=ttl, col sep=comma, mark=] {pdb.csv};
\end{axis}
\end{tikzpicture}
\caption{Number of protein structures in the Protein Data Bank (PDB) between 1991 and 2015.}
\end{figure}
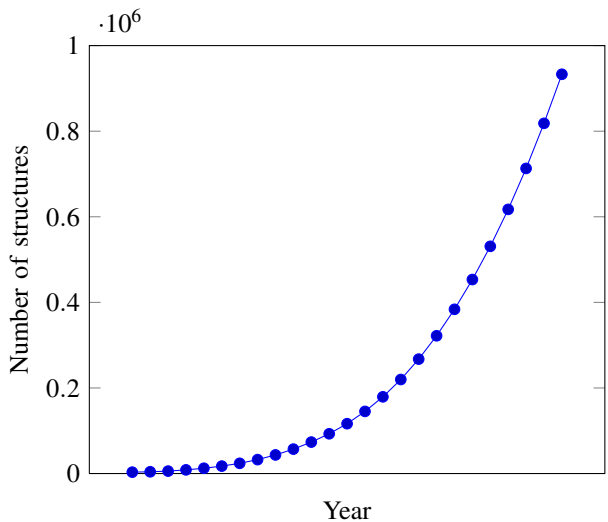
One method for the identification of structural neighbors is structural alignment, which attempts to discover similarity between two multiple protein structures based on their three-dimensional conformation. Performing structural alignment between the subject protein and all other proteins in the Protein Data Bank (PDB) via a structural alignment tool such as Protein Structure Comparison by Alignment of Distance Matrices (DALI) [15], Flexible Structure Alignment by Chaining Aligned Fragment Pairs Allowing Twists (FATCAT) [16], or Multiple Alignment with Translations and Twists (MATT) [17] can elucidate the discovery of the function of a protein of known structure but unknown function.\\\\
\indent Many methods evaluate similarity between structures of proteins via the comparison of $\alpha$- or $\beta$-carbon coordinates of amino acids as a representation for the subject proteins. After the amino acids of subject proteins are paired, subject proteins are superimposed and the original similarity function is modified to obtain a final solution, often optimizing the distances between already paired amino acids while preserving the three-dimensional conformations of all subject proteins. The final solution is always quasi-optimal as the aforementioned problem has been proven, in principle, to be NP-hard with no exact solution [18]. Thus, it is possible to see the vast differences of heuristics and scoring criteria, as noted by Mayr et al. (2007), in all protein structure alignment methods, which diversifies the so-called "optimal" solution for the structural alignment of a certain set of subject proteins [19].
One example of the heuristics used in alignment methods is the alternating approach to macromolecular docking: Many existing methods [15, 20 - 23] consider the subject proteins as rigid bodies, and attempt to minimize the deviation between the mapping of identified structures in all subject proteins while maximizing the number of mappings, describing the cumulative success in superimposition in terms of Root Mean Square Deviation (RMSD):
\begin{equation}
\mathrm{RMSD}=\sqrt{\frac{1}{N}\sum_{i=1}^N\delta_i^2}
\end{equation}
where $\delta$ is the distance between $N$ pairs of $\alpha$- or $\beta$-carbon atoms.\\\\
\indent Menke et al. (2008) showed that introducing flexibility to structural alignment would enable to align better with particularly marginal three-dimensional states (the three-dimensional folding of a protein changes accordingly to the number, location, or character of ligands attached to it), which cannot be achieved through rigid-body structural alignment. Performing structural alignment while allowing a flexible-body conformational state with twists and translations also facilitates the management of pairwise distortions outside the stable backbone [17]. Menke et al. (2008) also provided a pairwise structural alignment tool, namely Multiple Alignment with Translations and Twists (MATT), to demonstrate flexible-body structural alignment. Nevertheless, flexible structural alignment is computationally more exhaustive than rigid-body structural alignment: Menke et al. (2008) reported that Multiple Alignment with Translations and Twists (MATT) runs in $\mathcal{O}$($k^2n^3$ log $n$) time [17] compared to $\mathcal{O}$($k^2n^3$) during rigid-body structural alignment, reported by Konagurthu et al. (2006) [24], where $k$ is the number of protein structures being aligned, and $n$ is the primary structure length of the longest protein included in the alignment [17]. It is important to note that for small values of $n$, the discrepancy in running time is infinitesimal; but the aforementioned increase in protein structures in the PDB and the discovery of proteins of unknown function demand more exhaustive alignments, which in turn makes the rigid-body approach more preferable.\\\\
\indent
Looking at both commercialized and non-commercialized (open-source) pairwise and multiple structural alignment methods available, we report an arising problem in the daily use of these tools: Given the need for better methods to detect structural neighbors to a newly-discovered protein, the current methods do not offer a starting point to the user in terms of proteins of similar structure. Current methods take at least two protein structures as input, rendering the user optionless if the user has an unknown protein to which they would like to find structurally similar other proteins. In such a case, an exhaustive search through the PDB is optimal, as demonstrated by Holm et al. (1993) [15], and Godzik et al. (2004) [16]. Given the high-order time complexity of structural alignment, the running time of a full database search for a single query protein is too long to detect structurally similar proteins real-time (an average running time of 20 minutes up to an hour for a single query [25]).  Protein structure database search methods available also assume a non-redundant library, neglecting the possibility of finding interesting structural and functional variability in newly discovered structures [26]. Furthermore, these techniques can only be executed for annotated structures in the PDB, which hinders the potential use of structural alignment in determining the function of an unannotated protein that do not yet have a structural neighbor in the library. None of the PDB search methods available today do a complete and exhaustive search of the PDB, which is computationally expensive and demanding as it requires naive $\mathcal{O}(n)$ and $\mathcal{O}(n^2)$ comparisons; Furthermore, to our knowledge, no protein structure search and alignment tool that produces statistically significant flexible-body structural alignments using the database hits currently exists.\\\\
\indent
The aim of this study is establish a novel, robust, and accurate framework for protein structure database search and multiple structural alignment, through which we are able to search the entire Protein Data Bank (PDB) to find structural homologs of an input protein, and conduct pairwise structural alignment with the nearest structural neighbors allowing flexibility in shape and sort the resultant structural neighbors in descending order of statistical significance. We also present a web-based database search and structural alignment tool with a dynamic user interface to actualize this approach. We build upon some techniques proposed by Budowski-Tal et al. (2009) [27] and Yu et al. (2015) [28], while genuinely optimizing both the database search and the structural alignment.\\\\
\indent We approach the problem of searching the entire Protein Data Bank in an efficient way using the "filter and refine" method [29], where a computationally inexpensive search is conducted to choose a small set of potential structural neighbors, followed by a more exhaustive and demanding alignment technique. Usually via the representation of structures as  vectors in a $k$-dimensional hyperspace, techniques to search the PDB prove to be faster than raw structural alignment methods: Namely, Choi et al. (2004) use distance matrices inside the structure as a vector representation of frequencies, and quantize similarity by computing the distance between each vector (LFF) [30]; Rogen et al. (2003) utilize and adapt knot theory in algebraic topology and homotopy theory to engineer the Scaled Gauss Metric (SGM), representing structures as a vector of 30 topological quantities [31]; various techniques used a finite, orderly string of fragments, and consider the comparison and alignment of these vectors as a channel to detect structural similarity [32 - 34]. Although these filter methods easily outperform single-channel rigid structural alignment in running time, they report accuracy benchmark classifications that easily favor using full structural alignment over filter and refine methods: LFF, SGM, and PRIDE2 by Gaspari et al. (2005) [34] report classification accuracies of 68.7\%, 69.1\%, and 48.4\% respectively, benchmarked to the SCOP protein classification database [35], which are easily outperformed by full (both rigid-body and flexible) structural alignment tools. We take a filter-and-refine method initially proposed by Budowski-Tal et al. (2009) [27], which represents objects as an unordered collections of local structural features, where the query is described as a vector of the occurrences of the structural motifs in the query protein backbone, as a starting point for a quick and accurate structural neighbor retrieval method. \\\\ 
\indent 
Our best filter method easily outperforms other filter-and-refine approaches, and, more importantly, full-body structural aligners such as DALI [15] and FATCAT [16]; it can potentially be used to both efficiently, accurately, and rapidly identify good candidates of structural neighbors, and generate a set of potential structural neighbors for a protein with known structure and unknown function.
\section{APPROACH}
\subsection{Parameter optimization.}
We elaborate on the unordered vector space approach to protein structure similarity search proposed by Budowski-Tal et al. (2009), FragBag [27], which expresses proteins as a collection of frequencies of structural features with no specific order. FragBag utilizes a non-redundant library of structural fragments and motifs within proteins in the Protein Data Bank to collectively compute how similar two query proteins are, where the two query proteins are described as a collection of its contiguous structural fragments that overlap. The protein is then simply represented as a "bag-of-fragments", which is a $k$-dimensional vector that holds the frequency of each contiguous overlapping structural fragment as a separate entry within the vector. 
\begin{figure}[H]
\begin{subfigure}[H]{\columnwidth}
\centering
\begin{tikzpicture}
\node[anchor=south west,inner sep=0] (image) at (0,0)
{\includegraphics[width=0.6\textwidth]{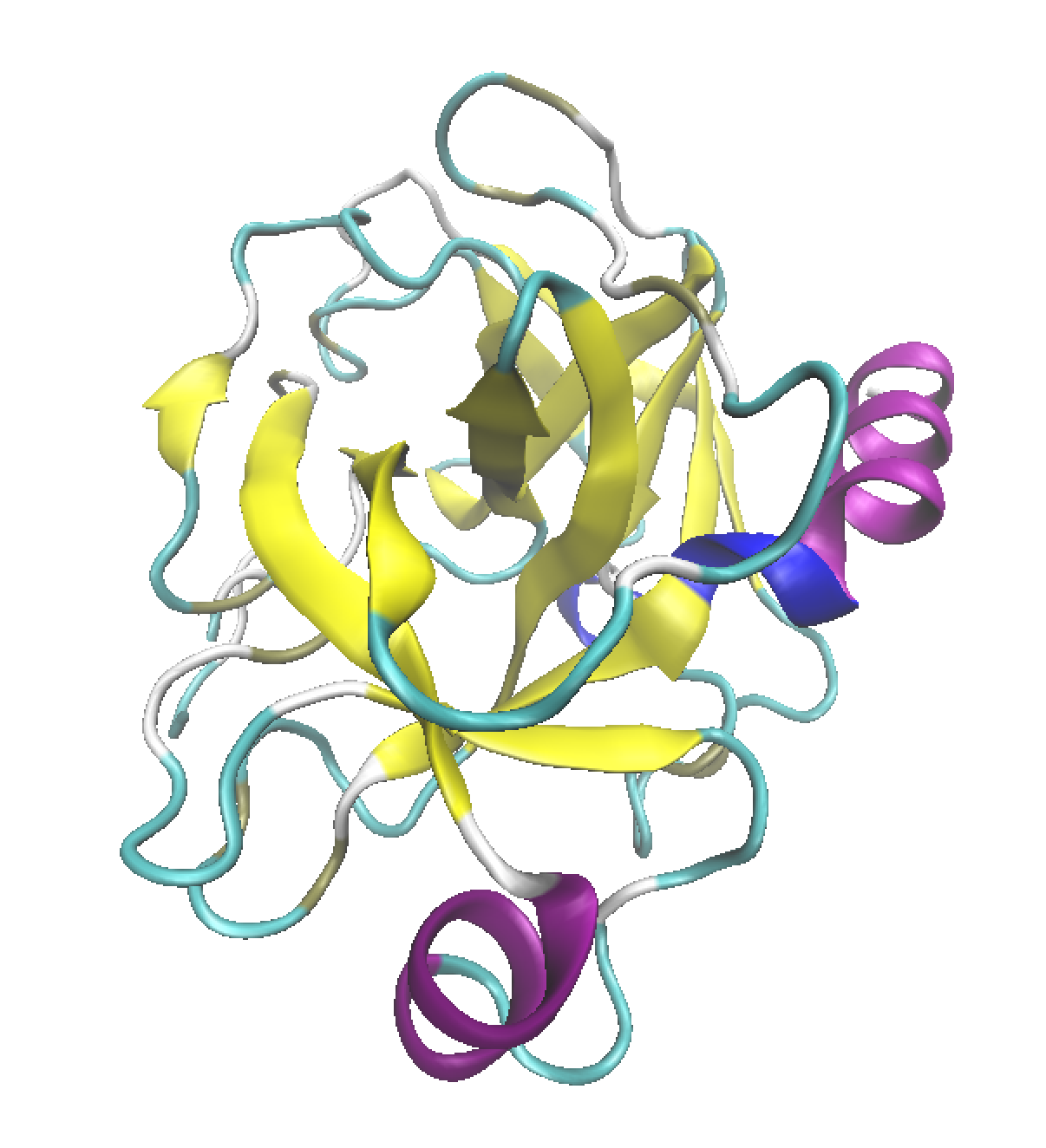}};
    \begin{scope}[x={(image.south east)},y={(image.north west)}]
        \draw[-latex] (0.87,0.7) -- +(0.1in,0.5in)node[anchor=west] {$f_{2}$};
        \draw[-latex] (0.58,0.6) -- +(0.8in,-0.7in)node[anchor=west] {$f_{4}$};
        \draw[-latex] (0.29,0.5) -- +(-0.6in,-0.15in)node[anchor=east] {$f_{3}$};
          \draw[-latex] (0.40,0.1) -- +(-0.6in,-0.1in)node[anchor=east] {$f_{6}$};
        \draw[-latex] (0.18,0.66) -- +(-0.35in,0.55in)node[anchor=east] {$f_{1}$};
        \draw[-latex] (0.5,0.38) -- +(0.7in,-0.7in)node[anchor=north] {$f_{5}$};
    \end{scope}
\end{tikzpicture}
\end{subfigure}
\begin{subfigure}[H]{\columnwidth}
\setlength{\unitlength}{0.8cm}
\begin{picture}(8,1.6)
\thicklines
\centering
\put(2,0.8){$\vec{v_{p}} = [s(f_{1}), s(f_{2}), s(f_{3}), s(f_{4}),s(f_{5}), s(f_{6})]$}
\end{picture}
\end{subfigure}
\caption{A simplified FragBag representation of porcine $\alpha$-trypsin (1AKS). After each unique contiguous backbone fragment of the query protein is identified, the occurrences $s$ of each segment is counted and put as entries in a $k$-dimensional query vector. For porcine $\alpha$-trypsin (1AKS), the resultant FragBag vector $\vec{v_{AKS}}$ would have six non-zero dimensions: [3, 1, 3, 6, 11, 1]. Note that all 400(11) library FragBag vectors put in a high-dimensional vector space representative of the protein structure dataset $S$ are 400-dimensional; yet, only the structures that are present in the backbone have a non-zero occurrence $s$.}
\label{fig:figure}
\end{figure}
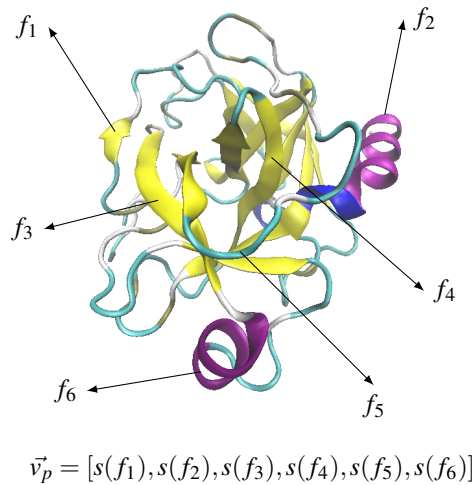
FragBag ultimately notes two benefits to representing a query protein in terms of its backbone segments:
\begin{itemize}
\item To implement a fast and efficient estimation of full database search of the Protein Data Bank;
\item For structure predictions where there is partial (fragment-wise), albeit no complete structural similarity.
\end{itemize}
Both of these aspects are favored in any approach to structural similarity; yet, we provide conceptual and methodological motivation to select FragBag as a starting point over other filter-and-refine methods: Namely, during the task of similarity search for structural neighbors in the high-dimensional vector space $S$ constructed by representative data points, as we widen our search radius, the vector space tends to not drastically increase, showing point density. This is to be elaborated later, particularly in II.B.\\\\
\indent
FragBag remains to be the gold standard in filter-and-refine methods mimicking full-body structural aligners, identifying structural neighbors on a par with some state-of-the-art full-body structural aligners within the 400(11) non-redundant structural fragment library. Budowski-Tal et al. (2009) reports robustly generated results from similar rankings with varying definitions of structural similarity, and report a cumulative success of identifying over 75\% of true positive structural neighbors benchmarked to the SCOP classification database [27] in all distance metrics benchmarked, with a maximum score of 89\% within the 400(11) non-redundant library.\\

\subsection{Mechanics of entropy-based search.} 
We generalize and justify a novel method that can be applied to not just similarity search within potential structural neighbors of a query protein but any kind of exhaustive search of -omics data. We approach the task of similarity search within the protein structure dataset by defining search performance in terms of the difference in novelty between new data and already existing data, \textit{entropy}, hence the name "entropy-based clustering". This approach enables the re-construction of the protein structure dataset, so that both the amount of time and space that the task of similarity search requires are linearly proportional to the entropy of the dataset, thus sublinearly proportional to the increase in size of the dataset itself. \\\\
\indent
We define two key elements of a dataset that will be vital to our approach: \textit{metric entropy} and \textit{fractal dimension}. We provide rigorous definitions for both of the concepts in Appendix A, but we are able to succinctly define metric entropy as describing the amount of dissimilarity a database appears to provide within itself, and fractal dimension as describing the relation between the number of clusters needed to cover all unique data points in a database and the respective radii of the clusters (note that the concept of metric entropy is different from that of a distance metric, which is a measure of distance applicable to any database). Via the analysis and generalization of these two elements of any database, we show that if the database in question (in our case, the protein structure database) appears to exhibit asymptotically low metric entropy and fractal dimension, the method outperforms naive full-body structural alignments and heuristically optimized filter-and-refine methods. The main advantage of optimizing entropy-based clustering and similarity search using metric entropy and fractal dimension is that it allows for a mathematical, instead of an experimental, evaluation of the approach in terms of  efficiency and accuracy. We also show that the entropy-based clustering approach results in zero loss in sensitivity.\\\\
\indent
The entropy-based similarity search is a quasi-$k$-means clustering algorithm, followed by an $n$-step hierarchical search that consists of four main steps.
\begin{enumerate}
\item Exhaustively analyze the database entries and define a high-dimensional vector space $S$, mapping each unique database entry onto unique points in this space $S$.
\item Use this space $S$ and a quantized measure of similarity
to group data points into clusters.
\item To search
for potential candidates, perform an initial search to identify the clusters that could possibly contain similar data points to the query.
\item Do a search within these clusters to find the closest data points to the query.
\end{enumerate}
We support this initial framework of hierarchical search by providing a conceptual supplemental rationale. We verbally approximate entropy as the vector distance between data points in this high-dimensional vector space $S$ (protein structures represented as vectors); hence, if $S$ exhibits low entropy, data points added to $S$ tend to not be distant from points already existing in $S$. We quantize the distance between data points in $S$ using generic distance metrics: Namely, Euclidean and cosine distance:\\\\
\indent For given queries $p = (p_{1}, p_{2}, ..., p_{n})$ and $q = (q_{1}, q_{2}, ..., q_{n})$, the Euclidean distance is:
\begin{equation}
d_{e}(p, q)=\sqrt{\sum_{i=1}^{n} (q_{i} - p_{i})^2}
\end{equation}
and the Cosine distance is:
\begin{equation}
d_{c}(p, q)= \frac{ \sum\limits_{i=1}^{n}{p_i  q_i} }{ \sqrt{\sum\limits_{i=1}^{n}{p_i^2}}  \sqrt{\sum\limits_{i=1}^{n}{q_i^2}}}
\end{equation}We experimentally evaluate both metrics in our process of creating a hierarchical clustering framework later.  \\\\
\indent Many studies defined the Protein Data Bank to be highly redundant [36 - 40], but it is vital that we define what being redundant for the protein structure data set signifies. Smith et al. (2015) posited that many of the data points of protein structures may be exact duplicates; that is, they may appear to have the same construction of representative vectors after the filter-and-refine method. This case can easily be solved by using a non-redundant library of proteins, as demonstrated by Ye et al. (2004) [16]. Maybe the representative data points exist in only a minimal number of dimensions; namely, showing low-dimensionality. If the dimension of the high-dimensional vector space is low enough, it can be divided into countable units, which would then ameliorate the running time for similarity searches. Although it is important to note that for datasets where empty space fills the vector space so that data points are sparsely distributed, many empty cells will be included in the search, significantly increasing running time. Furthermore, Yu et al. (2015) noted that many biological datasets do not reside in low-dimensions; instead, they arise from a highly chaotic high-dimensional vector space, properly characterized as a "tree of life" [28]. In these datasets, local low-dimensionality can be observed while the whole vector space is high-dimensional. This local low-dimensionality can be harnessed to search through specific regions of low-dimension with entropy-based similarity search.
\begin{figure}[H]
      \centering
     \includegraphics[width=0.5\textwidth]{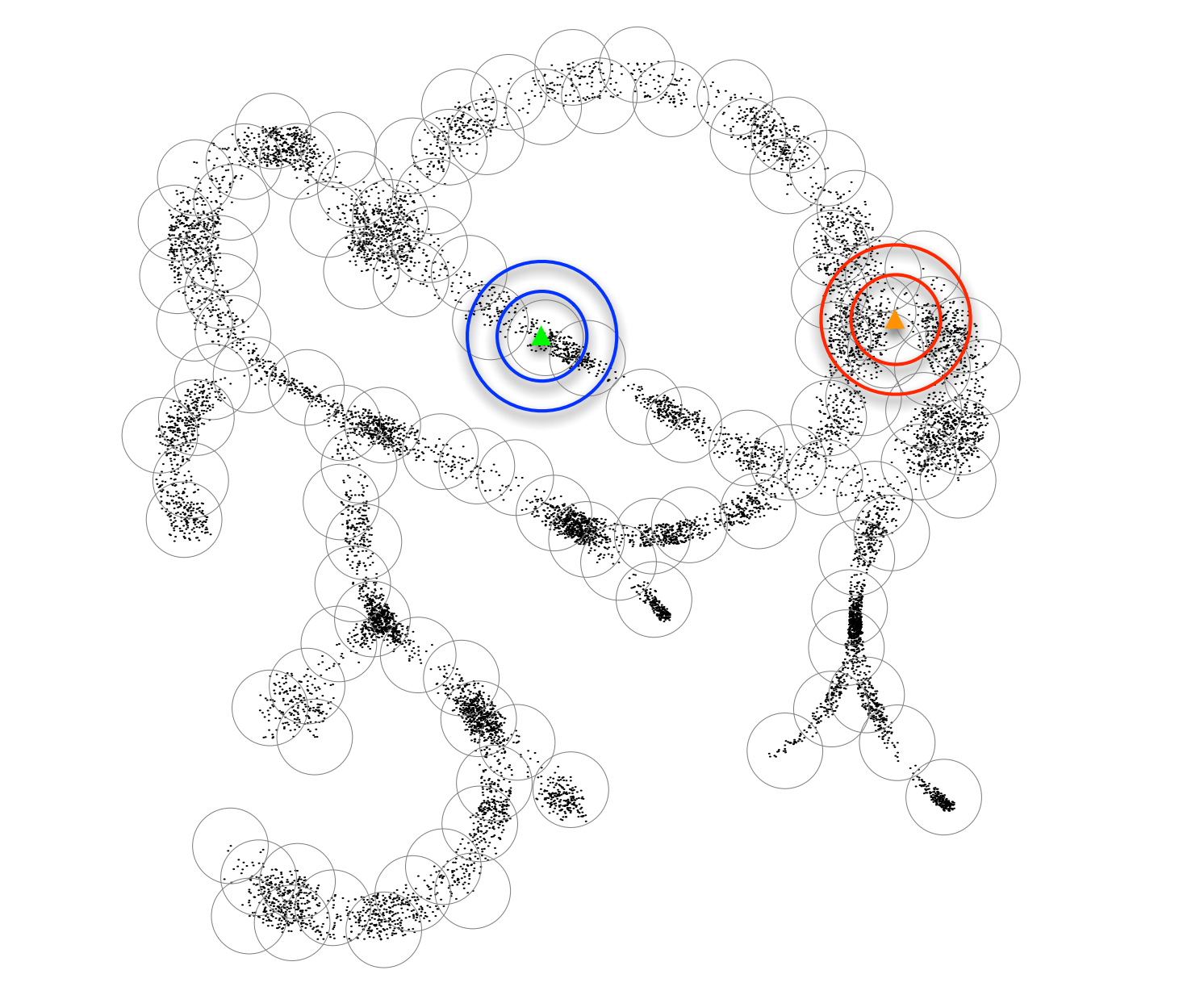}
      \caption{Depiction of representative data points in an arbitrary high-dimensional vector space exhibiting local low-dimensionality and global high-dimensionality by Yu et al. (2015). Reprinted from [28].}
      \label{figurelabel}
   \end{figure}
As seen in Figure 2, -omics data reside in high-dimensional vector spaces, but given a coarse scale to observe in this high-dimensional space, the density and the distribution of the representative data points are almost unidimensional. It is important to note that, based on the aforementioned succinct definition of fractal dimension, the high-dimensional vector space $S$ exhibits low local fractal dimension within the blue circles around the green query, and high fractal dimension within the red circles around the orange query. The blue circles around
the green query point illustrate low fractal dimension: the larger-radius circle
contains only linearly more points than the smaller one, rather than
exponentially more. In contrast, the red circles around the orange query
point illustrate higher local fractal dimension.\\\\
\indent
Approaching this juxtaposition of different amounts of dimensionality in the same database, we may be able to exploit the local and global discrepancy: Using a wide enough search radius, the entropy-based approach assumes a unidimensional space, and as the search radius is reduced, we can gradually start to consider the branches of high-dimensionality. For this high-dimensional vector space $S$, we assume a coverage with spheres with radius $r_{c}$, where $r_{c}$ is equals the low-dimensional branch width in question, and
\begin{equation}
E_{r}=\sum_{i=1}^kS_i
\end{equation}
where $E_{r}$ is the metric entropy of the high-dimensional vector space $S$ and $S_i$ is the number of spheres needed to cover $S$.\\
We note that by the use of a spherical cluster, we assume that the points on this high-dimensional vector space $S$ are very close, thus can be encoded in terms of one another, which is in parallel with the aforementioned redundancy in the Protein Data Bank. Thus, using the triangle inequality, we are able to search all points within $r$ by only looking at adjacent spheres with cluster centers within $r + r_c$ of the query. For a wide search radius, we see that the spheres needed to cover $S$ extend along a minimal number of dimensions: As the radius decreases, the depth of the sphere matters. We call this property of local low-dimensionality the fractal dimension $d$ of the high-dimensional vector space $S$ at the radius scale $r_{c}$. Local fractal dimension $d$ in $S$ is computed via the increase between radii $r_1$ and $r_2$ over the points added to $S_{r_1}$, let us call these new points $n_2$ and previous points $n_1$, by the increase in radii. The local fractal dimension is then simply:
\begin{equation}
d = \frac{\log(n2/n1)}{\log(r2/r1)}
\end{equation}
Intuitively, we note that when $d=1$, the number of points added to the sphere $S_{r_1}$ is linear to the increase in radii, thus, our approach is maximally efficient when the fractal dimension $d$ and metric entropy $k$ is minimal. 
Yu et al. (2015) reports $2<d<3$ for average local fractal dimension of FragBag vectors [28], thus providing a rationale to investigate our approach over the protein structure high-dimensional vector space. When we search in a wider radius around a query, the number of points added to the spherical cluster covering the points around the query within the radius grows exponentially with the fractal dimension; implying that this growth will not hinder the efficiency provided
by an entropy-based search. We provide a rigorous investigation in Appendix A: Theoretical Foundations that given a high-dimensional vector space $S$ with a fractal dimension $d$, an entropy $k$, and an initial search radius $r_c$, the time complexity $T(n)$ of the entropy-based similarity search is 
\begin{equation}
T(n)=\mathcal{O}(k+D_{C}(q, r)(\frac{r + 2r_{c}}{r})^d)
\end{equation}
which is asymptotically linear to $k$ with minimal values of output size and fractal dimension.
\subsection{Clustering algorithm.}
\indent 
When the problem involves comparing properties of a data point to one another, the most
straightforward way to reduce computational expense is to cluster already existing data
points in the database prior to the search.  To reduce the running time for a query as much as possible, we used
hierarchical divisive clustering as it offers a complexity reduction from $\mathcal{O}(n)$ to $\mathcal{O}($log $n)$ for
query search. The goal of our approach is to balance the computational expense on
both sides; for a reduction from $\mathcal{O}(n)$ to $\mathcal{O}($log $n)$ for query search compensates for $\mathcal{O}(2^n)$ on
our side since the user can withstand much less computational burden.\\\\
\begin{figure*}[ht]
  \centering
    \begin{subfigure}[H]{0.5\textwidth}
        \centering
        \begin{tikzpicture}
 \centering
\begin{axis}[
legend style={at={(0.5,-0.1)},anchor=north},
ylabel near ticks,
xlabel near ticks,
xmin=-12, 
xmax=108, 
ymin=7, 
ymax=105,
ticks=none,
yticklabels={,,},
xticklabels={,,},
]
\addplot table [only marks, x=x, y=y, col sep=comma, mark=*] {co.csv};
\end{axis}
\end{tikzpicture}
        \caption{}
    \end{subfigure}%
    ~ 
    \begin{subfigure}[H]{0.5\textwidth}
        \centering
        \begin{tikzpicture}
 \centering
\begin{axis}[
legend style={at={(0.5,-0.1)},anchor=north},
ylabel near ticks,
xlabel near ticks,
xmin=-12, 
xmax=108, 
ymin=7, 
ymax=105,
ticks=none,
yticklabels={,,},
xticklabels={,,},
]
\addplot table [only marks, black, x=x, y=y, col sep=comma, mark=* ] {co.csv};
\draw[color=red, thick] (axis cs:46,33) circle (20);
\addplot[only marks, red, mark size=3, mark=*] coordinates {(46, 33)};
 \addplot[only marks, red, mark=*] coordinates {
(46,33)
(47,14)
(50,50)
(51,50)
(53,25)
(53,34)
(59,29)
(52,17)
(27,36)
(31,32)
(32,43)
(34,35)
(38,30)
(42,42)
(45,19)

};
\end{axis}
\end{tikzpicture}
        \caption{}
    \end{subfigure}%
        \par\bigskip 
    \begin{subfigure}[H]{0.5\textwidth}
        \centering
        \begin{tikzpicture}
 \centering
\begin{axis}[
legend style={at={(0.5,-0.1)},anchor=north},
ylabel near ticks,
xlabel near ticks,
xmin=-12, 
xmax=108, 
ymin=7, 
ymax=105,
ticks=none,
yticklabels={,,},
xticklabels={,,},
]
\addplot table [only marks, x=x, y=y, col sep=comma, mark=* ] {co.csv};
\draw[color=red, thick] (axis cs:46,33) circle (20);
\addplot[only marks, red, mark size=3, mark=*] coordinates {(46, 33)};
 \addplot[only marks, red, mark=*] coordinates {
(46,33)
(47,14)
(50,50)
(51,50)
(53,25)
(53,34)
(59,29)
(52,17)
(27,36)
(31,32)
(32,43)
(34,35)
(38,30)
(42,42)
(45,19)
};
\draw[color=green, thick] (axis cs:32,56) circle (20);
\addplot[only marks, green, mark size=3, mark=*] coordinates {(32, 56)};
 \addplot[only marks, green, mark=*] coordinates {
(18,65)
(22,69)
(25,70)
(37,71)
(17,51)
(16,88)
(22,93)
(32,56)
(34,58)
(34,55)
(14,66)
(14,69)
(26,40)
(17,43)
(41,54)
(41,62)
(17,43)
};
\draw[color=yellow, thick] (axis cs:87,20) circle (20);
\addplot[only marks, yellow, mark size=3, mark=*] coordinates {(87, 20)};
 \addplot[only marks, yellow, mark=*] coordinates {
(82,4)
(85,5)
(89,5)
(85,11)
(84,14)
(88,14)
(86,17)
(87,20)
(75,6)
(75,16)
(70,13)
(76,23)
(71,31)
(70,25)
(82,33)
(97,33)
(99,33)
(100,32)
(100,29)
(100,19)
(99,22)
};
\draw[color=cyan, thick] (axis cs:11,80) circle (20);
\addplot[only marks, cyan, mark size=3, mark=*] coordinates {(11, 80)};
 \addplot[only marks, cyan, mark=*] coordinates {
(3,96)
(6,96)
(11,100)
(10,89)
(12,89)
(16,88)
(22,93)
(11,80)
(6,76)
(11,64)
(14,66)
(14,69)
(24,78)
(27,79)
};
\draw[color=purple, thick] (axis cs:53,85) circle (20);
\addplot[only marks, purple, mark size=3, mark=*] coordinates {(53, 85)};
 \addplot[only marks, purple, mark=*] coordinates {
(51,73)
(61,72)
(54,81)
(53,85)
(57,80)
(52,88)
(36,94)
(44,94)
(41,93)
(41,91)
(69,96)
};
\draw[color=pink, thick] (axis cs:83,64) circle (20);
\addplot[only marks, pink, mark size=3, mark=*] coordinates {(83, 64)};
 \addplot[only marks, pink, mark=*] coordinates {
(92,82)
(93,74)
(97,69)
(96,62)
(92,60)
(95,51)
(83,64)
(86,67)
(96,49)
(66,68)
(70,71)
(78,70)
(68,54)
(75,51)
};
\draw[color=orange, thick] (axis cs:9,31) circle (20);
\addplot[only marks, orange, mark size=3, mark=] coordinates {(9, 31)};
 \addplot[only marks, orange, mark=] coordinates {
(1,46)
(4,45)
(10,45)
(12,42)
(9,49)
(9,31)
(2,14)
(9,39)
(18,18)
(21,19)
(15,33)
(19,26)
};
\end{axis}
\end{tikzpicture}
        \caption{}
    \end{subfigure}%
    ~ 
    \begin{subfigure}[H]{0.5\textwidth}
        \centering
         \begin{tikzpicture}
 \centering
\begin{axis}[
legend style={at={(0.5,-0.1)},anchor=north},
ylabel near ticks,
xlabel near ticks,
xmin=21, 
xmax=72, 
ymin=12, 
ymax=54,
ticks=none,
yticklabels={,,},
xticklabels={,,},
]
\addplot table [only marks, x=x, y=y, col sep=comma, mark=* ] {co.csv};
\draw[color=red, thick] (axis cs:46,33) circle (200);
\draw[color=black, thick] (axis cs:53,34) circle (100);
\addplot[only marks, red, mark size=3, mark=*] coordinates {(46, 33)};
 \addplot[only marks, red, mark=*] coordinates {
(46,33)
(47,14)
(50,50)
(51,50)
(53,25)
(53,34)
(59,29)
(52,17)
(27,36)
(31,32)
(32,43)
(34,35)
(38,30)
(42,42)
(45,19)
};
\draw[color=green, thick] (axis cs:32,56) circle (200);
\addplot[only marks, green, mark size=3, mark=*] coordinates {(32, 56)};
 \addplot[only marks, green, mark=*] coordinates {
(18,65)
(22,69)
(25,70)
(37,71)
(17,51)
(16,88)
(22,93)
(32,56)
(34,58)
(34,55)
(14,66)
(14,69)
(26,40)
(17,43)
(41,62)
(17,43)
};
\draw[color=yellow, thick] (axis cs:87,20) circle (200);
\addplot[only marks, yellow, mark size=3, mark=*] coordinates {(87, 20)};
 \addplot[only marks, yellow, mark=*] coordinates {
(82,4)
(85,5)
(89,5)
(85,11)
(84,14)
(88,14)
(86,17)
(87,20)
(75,6)
(75,16)
(70,13)
(76,23)
(71,31)
(70,25)
(82,33)
(97,33)
(99,33)
(100,32)
(100,29)
(100,19)
(99,22)
};
\draw[color=cyan, thick] (axis cs:11,80) circle (20);
\addplot[only marks, cyan, mark size=3, mark=*] coordinates {(11, 80)};
 \addplot[only marks, cyan, mark=*] coordinates {
(3,96)
(6,96)
(11,100)
(10,89)
(12,89)
(16,88)
(22,93)
(11,80)
(6,76)
(11,64)
(14,66)
(14,69)
(24,78)
(27,79)
};
\draw[color=purple, thick] (axis cs:53,85) circle (20);
\addplot[only marks, purple, mark size=3, mark=*] coordinates {(53, 85)};
 \addplot[only marks, purple, mark=*] coordinates {
(51,73)
(61,72)
(54,81)
(53,85)
(57,80)
(52,88)
(36,94)
(44,94)
(41,93)
(41,91)
(69,96)
};
\draw[color=pink, thick] (axis cs:83,64) circle (20);
\addplot[only marks, pink, mark size=3, mark=*] coordinates {(83, 64)};
 \addplot[only marks, pink, mark=*] coordinates {
(92,82)
(93,74)
(97,69)
(96,62)
(92,60)
(95,51)
(83,64)
(86,67)
(96,49)
(66,68)
(70,71)
(78,70)
(75,51)
};
\draw[color=orange, thick] (axis cs:9,31) circle (20);
\addplot[only marks, orange, mark size=3, mark=] coordinates {(9, 31)};
 \addplot[only marks, orange, mark=] coordinates {
(1,46)
(4,45)
(10,45)
(12,42)
(9,49)
(9,31)
(2,14)
(9,39)
(18,18)
(15,33)
(19,26)
};
\end{axis}
\end{tikzpicture}
        \caption{}
    \end{subfigure}
\caption{A simplified visual representation of the entropy-based flat divisive hierarchical clustering algorithm utilized. Each protein is represented by a unique data point in a high dimensional vector space (a). A random cluster center $k$ is chosen and the points within a radius $r$ are put in this cluster (b). Then, iteratively all points are clustered via the same approach, and this process repeats until all points are in a cluster (c). Once the first level of clustering is done, data points within a cluster are clustered with a smaller radius $r'$ (d). Note that the data points at the intersection of overlapping clusters are assigned based on the order of clustering: In (a), the red cluster is the first cluster formed, thus containing data points that are also in green and orange clusters.
}
\label{figurelabel}
   \end{figure*}
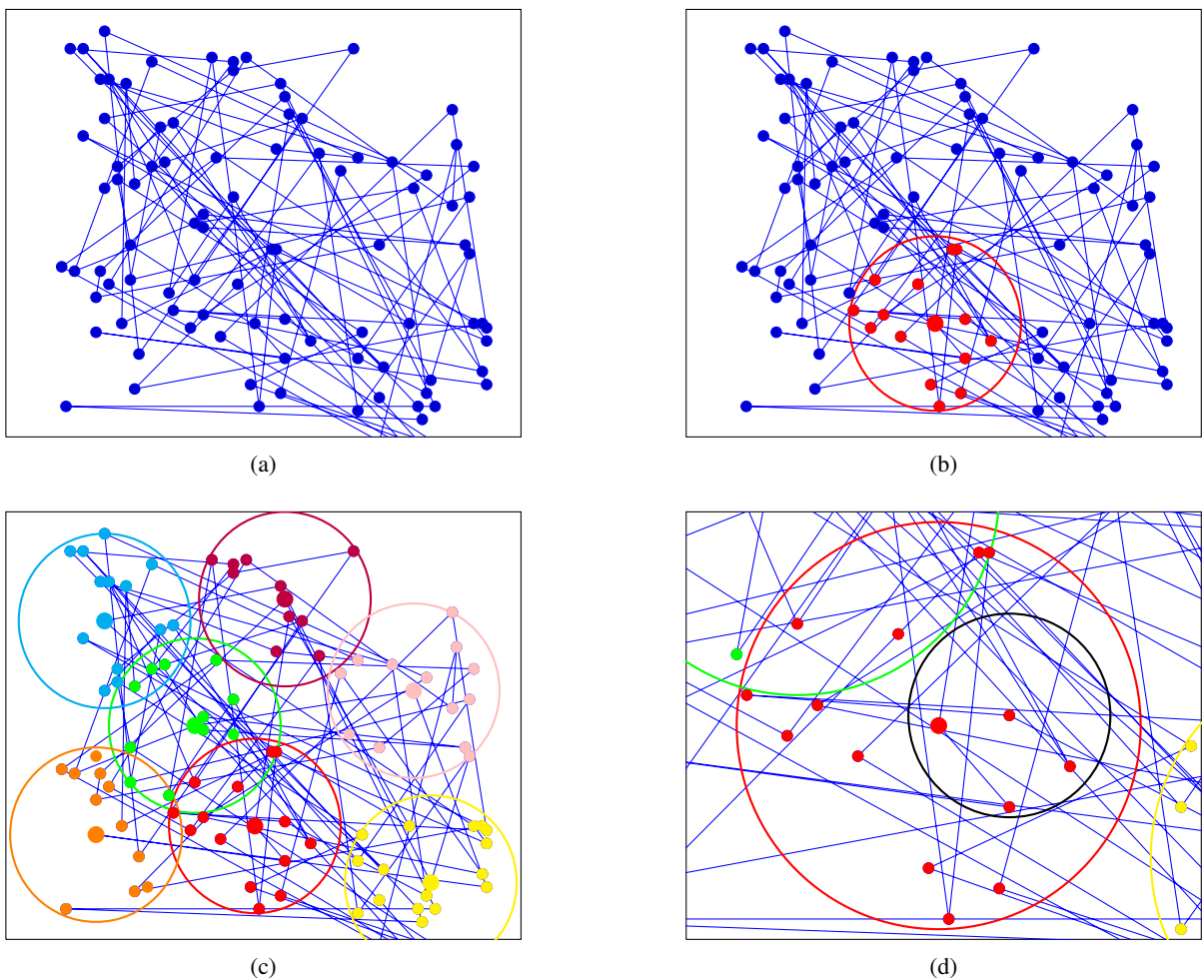
\indent We present a novel algorithm that relies on the hierarchical exhaustive logic that drives
every single data point to eventually be part of a cluster. Once the real-world objects (proteins)
are expressed as vectors and are put in a high-dimensional vector space $S$ (the whole
Protein Data Bank), an initial cluster center $k$ is randomly chosen in this high
dimensional space $S$. $k$ is a data point in $S$ itself. A user defined radius of $r$ defines a circular
grid in which all the points are considered a part of the cluster by the user. In other words, once
$k$ is chosen in $S$, the user decides to what extent should the data points be represented
by the cluster center $k$ by giving $r$ as an input.\\\\
\indent A data point $i$ is part of the cluster $C_{k}$
with the cluster center $k$, where the distance value between $i$ and $k$ $D_{ik} \leq r$. Thus, a cluster
is formed based on a user-defined $r$. Then, the cluster is temporarily removed
from $S$ and another randomly chosen cluster center $k'$ is investigated for data points with
distance less than $r$. This is repeated until every data point $i$ in the high dimensional vector
space $S$ is part of a cluster with a center $k$. It is irrelevant to project the amount of initial
clusters after input $r$ since it very much depends on the data set and the magnitude of $r$ and
does not affect the output since the only prerequisite for the next step is to make sure every
data point is part of a cluster.\\\\
\indent Once every data point in $S$ is a member of a cluster, the dummy timer $t = t + 1$, and
the user-defined radius $r$ is reduced by a user-defined function $R(r)$ where $t = t + 1 \rightarrow$
 $R(r) = r'$ for both the search and the clustering. Some options of $r$
is automatically given to the user for the query search, such as $\sqrt{r}$ and $r/2$, and users are able to define $R(r)$ as long as it outputs a value where $r' < r$. Within each cluster, another randomly assigned cluster center $k'$ and the decreased radius $r'$ is put in to find data points similar to $k'$ and include them in the new cluster $C_{k'}$ where now data point $j \in C_{k'}$ if $D_{jk'} \leq r$. Once this process is done, the dummy timer $t = t + 1.$
\begin{figure}[H]
\begin{algorithm}[H]
\begin{algorithmic}[1]
\Require $r \neq 0, r \in \mathbb{Z^+};$
\Ensure all points $n \in S$;
\State $\textit{k} \gets \text{random point in $S$;}$
\State $C_{k} \gets$ cluster with center $k$;
\State $D_{ik} \gets$ distance between $i$ and $k$;
\BState \emph{top}:
\If {$\exists C_{k} \mid n \in C_{k}$} \textbf{quit};
\Else
\BState \emph{loop}:
\If {$D_{nk} \leq r$} 
\State $n \in C_k$;
\EndIf
\State \textbf{goto} \emph{loop};
\State \textbf{close};
\EndIf
\State $r \gets r'$;
\State \textbf{hide} $C_k$;
\State \textbf{goto} \emph{top}.
\end{algorithmic}
\end{algorithm}
\caption{Pseudocode for the hierarchical clustering algorithm.}\label{euclid}
\end{figure}
\indent Once all the clustering within all existing clusters completes, the clustering continues
with radius $r$ getting reduced by both the user-defined function $R(r)$ and the dummy timer
$t = t + 1$. After each data point in the vector space $S$ is a cluster within itself, the process
stops. $t$ yields the number of levels of clustering that can depend on the data set and radius
$r$ and also, incidentally, the level of the deepest point (the point within the most number of
clusters). The algorithm assumes that the system is unsupervised in the sense that it will
always reach the static state where there is no clustering that needs to be done. To make the
algorithm more flexible, an optional user-defined operator $d$ is introduced where $d$ denotes
how deep the clustering should go. The algorithm quits when $d = t$.\\\\
\indent Ultimately, the high dimensional vector space $S$ will include at least $n$ clusters for $n$ data
points, and probably more. Yet, this final specificity is redundant as the only member
of the final cluster is the data point itself: When d is specified, the algorithm quits at $d-1$
to prevent redundant search of clusters $C_{n}$ where $n = 1$.
Once a query is submitted to the system, cluster centers are identified in a user-defined
radius $r$. Cluster centers which are in this circular grid of radius $r$ is only searched when
$t = t + 1$ for radius $r'$.

\section{RESULTS}
\subsection{Scaling behavior.}
To obtain an optimal distance metric to be used in our entropy-based search, we investigated the increase in speed between both Cosine and Euclidean distance functions. The resulting values imply that the acceleration across metrics depend largely on the initial cluster radius. We produced databases for initial cluster radii of {0.1, 0.2, 0.3, 0.4, and, 0.5} and 10, 20, 25, 50, and 100, for Cosine and Euclidean distance respectively, with a unit change of 0.01 (50 unique values of cluster radius in total). We ran similarity searches for 4 query proteins with varying size and number of structural neighbors: Porcine $\alpha$-trypsin (1AKS), flavodoxin (1AKR), HIV-1 reverse transcriptase (1FKO), and hemoglobin-A (1ASH). We used reduction functions $\sqrt{r}$ and $r/2$ for $R(r)$; yet, due to the minimal change in results, reported an average of the two functions. We ran each query protein 5 times through the entropy-based structural similarity search for each cluster radius and included the averages for each cluster radius as one data point in our investigation.\\\\
\indent We also report a comparison of clustering time of both Euclidean and Cosine distances with aforementioned unique cluster radii. We compared the average clustering time over the whole high-dimensional vector space $S$ across five trials for each unique cluster radius. We used all 20 threads on an Intel Xeon x5690 (12-core) while clustering, compared to only one during search.
\begin{table}[]
\centering
\caption{Clustering time using Cosine distance}
\label{my-label}
\begin{tabular}{@{}cccccc@{}}
\toprule
\textbf{Radius}              & 0.1   & 0.2   & 0.3  & 0.4  & 0.5  \\ \midrule
\textbf{Clustering time (s)} & 22546 & 13243 & 7645 & 5433 & 4087 \\ \bottomrule
\end{tabular}
\end{table}
\begin{table}[]
\centering
\caption{Clustering time using Euclidean distance}
\label{my-label}
\begin{tabular}{@{}cccccc@{}}
\toprule
\textbf{Radius}              & 10   & 20  & 25  & 50  & 100  \\ \midrule
\textbf{Clustering time (s)} & 24407 & 3154 & 679 & 204 & 87 \\ \bottomrule
\end{tabular}
\end{table}
\begin{figure*}
\begin{subfigure}{0.5\textwidth}
 \centering
 \begin{tikzpicture}
 \centering
\begin{axis}[
ylabel near ticks,
xlabel near ticks,
xmin=0, 
xmax=0.5, 
ymin=0, 
ymax=6,
xlabel={},
ylabel={}]
\addplot table [scatter, x=rad, y=euc, col sep=comma, mark=*]{res.csv};
\addplot table [scatter, x=rad, y=cos, col sep=comma, mark=*] {res.csv};
\end{axis}
\end{tikzpicture}
      \caption{porcine $\alpha$-trypsin (1AKS)}
      \label{figurelabel}
              \vspace*{6mm}
    \end{subfigure}%
    \quad
    \begin{subfigure}{0.5\textwidth}
     \centering
 \begin{tikzpicture}
 \centering
\begin{axis}[
ylabel near ticks,
xlabel near ticks,
xmin=0, 
xmax=0.5, 
ymin=0, 
ymax=6,
xlabel={},
ylabel={}]
\addplot table [scatter, x=rad, y=euc, col sep=comma, mark=*]{res2.csv};
\addplot table [scatter, x=rad, y=cos, col sep=comma, mark=*] {res2.csv};
\end{axis}
\end{tikzpicture}
      \caption{flavodoxin (1AKR)}
      \label{figurelabel}
                    \vspace*{6mm}
    \end{subfigure}
        \begin{subfigure}{0.5\textwidth}
         \centering
 \begin{tikzpicture}
 \centering
\begin{axis}[
ylabel near ticks,
xlabel near ticks,
xmin=0, 
xmax=0.5, 
ymin=0, 
ymax=9,
xlabel={},
ylabel={}]
\addplot table [scatter, x=rad, y=euc, col sep=comma, mark=*]{res3.csv};
\addplot table [scatter, x=rad, y=cos, col sep=comma, mark=*] {res3.csv};
\end{axis}
\end{tikzpicture}
      \caption{HIV-1 reverse transcriptase (1FKO)}
      \label{figurelabel}
    \end{subfigure}%
        \quad
    \begin{subfigure}{0.5\textwidth}
         \centering
 \begin{tikzpicture}
 \centering
\begin{axis}[
ylabel near ticks,
xlabel near ticks,
xmin=0, 
xmax=0.5, 
ymin=0, 
ymax=11,
xlabel={},
ylabel={}]
\addplot table [scatter, x=rad, y=euc, col sep=comma, mark=*]{res4.csv};
\addplot table [scatter, x=rad, y=cos, col sep=comma, mark=*] {res4.csv};
\end{axis}
\end{tikzpicture}
      \caption{Ascaris hemoglobin (1ASH)}
      \label{figurelabel}
    \end{subfigure}
    \caption{Comparative scaling behavior of two different distance metrics, Cosine (red) and Euclidean (blue) used in entropy-based clustering and search (x-axis denotes the initial cluster radii, and y denotes running time in seconds). It is important to note that across a cluster radius of 0.5, $d$ ultimately approaches zero, providing the motivation to investigate redundancy within the protein structure high-dimensional vector space (as the cluster radius increases, more points are introduced to the search, but the rate in which these points are introduced decreases rapidly, approaching zero).}
   \end{figure*}
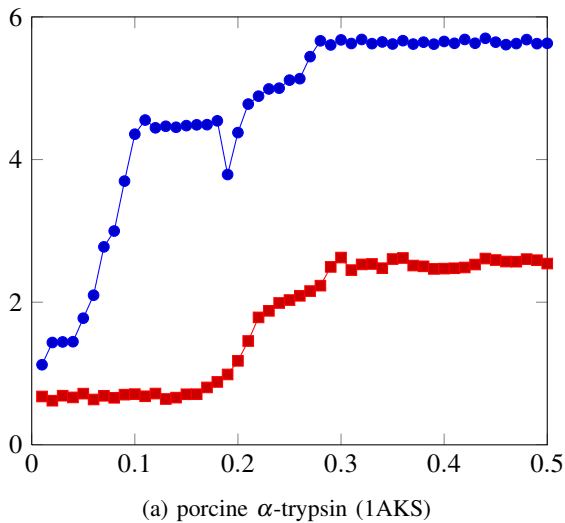
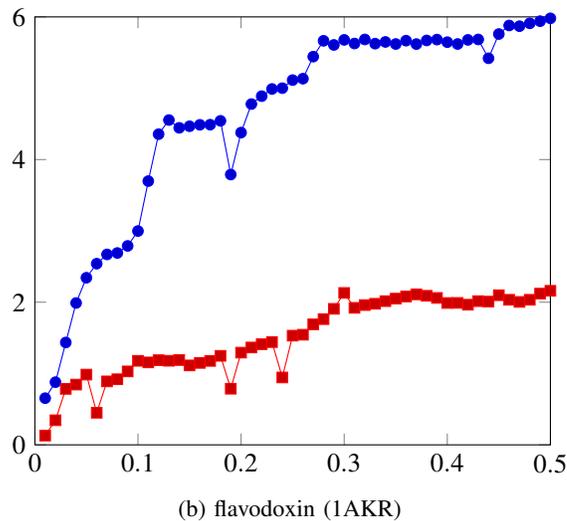
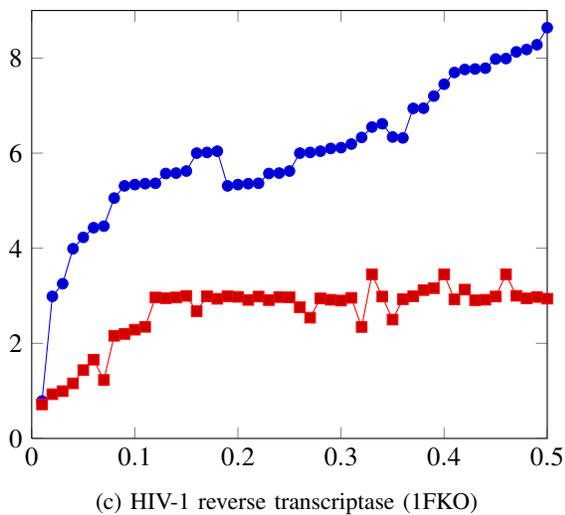
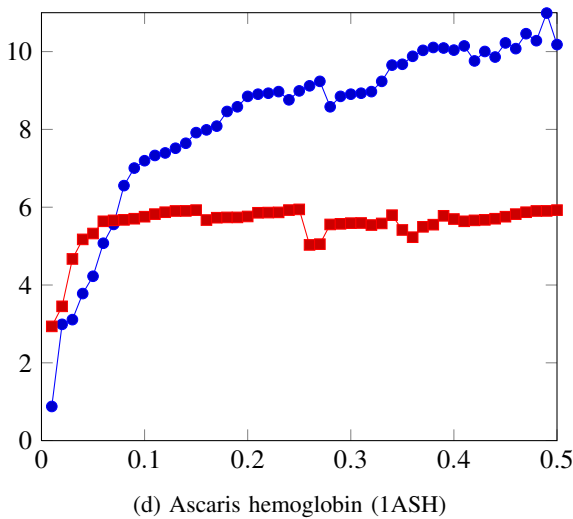
We have concluded that the running time for the task of similarity search in our entropy-based hierarchical clustering largely depends on the magnitude of the initial cluster radius $r_c$, and the number of structurally similar proteins that are around the query protein in the high-dimensional vector space $S$. As we demonstrate using the aforementioned 4 test query proteins with different sizes and degrees of membership in $S$, the acceleration across uniquely different values of $r_c$ decreases in different rates. We posit that this effect is most likely to be caused by the high-dimensional vector space $S$ having a lot of local density regions that are distinctly separated from other more uniformly distributed regions (being more "spiky" in largely populated, dense regions compared to peripheral regions). A protein that is a neighbor to many others in a densely populated region shows little to no acceleration with an increase in radius since a small increase covers a large part of the neighbors, destroying any acceleration. Modifying $r_c$ may enable the user to effectively search for proteins that are less likely to be a part of these aforementioned dense populations; yet, as $r_c$  increases, the running time increases proportionally.\\\\ 
\indent Based on our investigation of scaling behavior, we note that Cosine
distance generally yields better acceleration, and the Euclidean distance guarantees 100\% sensitivity via
Triangle Inequality [28]. Hence, we will be using Cosine distance as our speed benchmark candidate and Euclidean distance for our accuracy benchmark candidate.
\subsection{Benchmarking classifications}
We evaluated the accuracy and running time of our method by benchmarking the raw results to gold standards in protein structure similarity search; ideally, our method would outperform filter-and-refine methods (that are compromised in accuracy compared to full-body structural aligners) in speed, and outperform full-body structural aligners (which are compromised in speed compared to filter-and-refine methods) in accuracy. We have established the respective sets of gold-standards as being the most successful performers of both groups: FragBag [10], SGM [31], and TM-Align [41]; and DALI [15], FATCAT [16], and 3D-BLAST [42], for speed and accuracy benchmark classifications respectively. For the following results, we benchmarked our method, Esperite, along with other respective methods for both types of benchmark classifications.\\\\
\indent When comparing techniques, it is vital to realize that the monitoring of the relative performance of two methods rely heavily on not only the nature of the query and the database, but also the heuristics used in the technique that the subject method is being benchmarked to. We assume an independent classification database, the SCOP classification database [35] to distinctly separate our method from constructing derivations or improvements on other existing methods. We assume that these protein structure classification databases provide perfectly similar protein structural neighbors as the gold standard (no false positives or negatives). We use the SCOPe\_FAMILY group as the definition of closely related proteins, and the fold and super-family groups for distant proteins.
We utilize and evaluate receiver operating characteristic (ROC) curves as a binary computation of the performance of classifier methods: The curve depicts true positive rate (TPR) (number of SCOP structural neighbors of the query found) vs. the false positive rate (FPR) (number of SCOP structural non-neighbors of the query found) at various thresholds. To cumulatively present the accuracy of a method, we averaged the benchmarking results both across different queries and different classification methods within the classification database. We trained and evaluated the accuracy of Esperite across SCOPe\_FAMILY (2,623 domains in 903 superfamilies; 591 folds).
\\
\subsubsection{Search benchmarks.}
For benchmarking over searching for structural neighbors across the SCOP database, we plot the fraction coverage of the whole database versus the cumulative amount of false positives found, as a derivation to the traditional receiver operating characteristic (ROC). We plot the quasi-ROC curve for all SCOP subdatabases: for family, super-family, and fold classification levels.
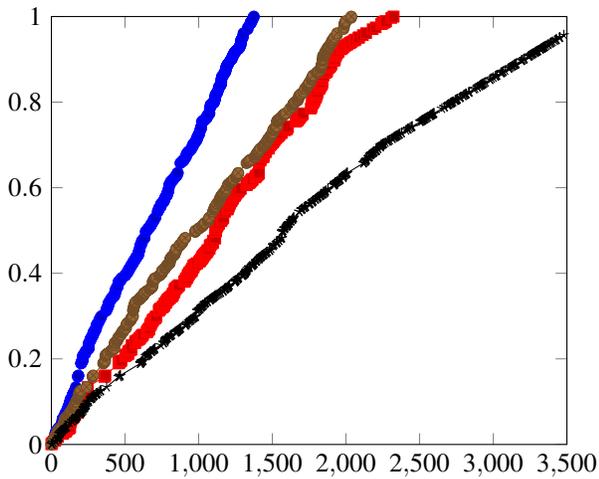
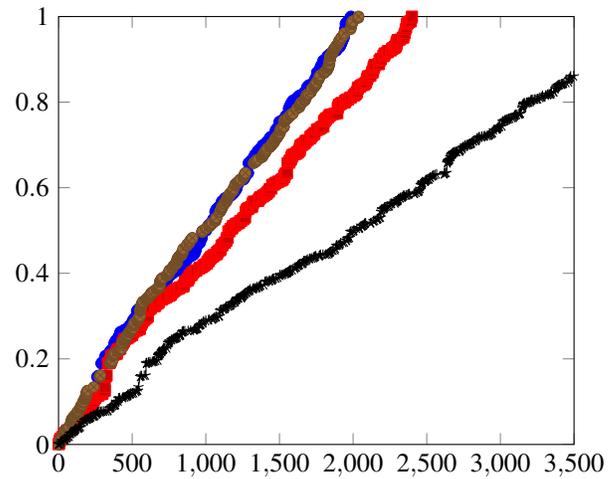
\begin{figure*}
\begin{subfigure}{0.5\textwidth}
 \centering
 \begin{tikzpicture}
 \centering
\begin{axis}[
ylabel near ticks,
xlabel near ticks,
xmin=0, 
xmax=3500, 
ymin=0, 
ymax=1,
xlabel={},
ylabel={}]
\addplot table [x=esp, y=fraction, col sep=comma, mark=*]{s_family.csv};
\addplot table [x=fatcat, y=fraction, col sep=comma, mark=*]{s_family.csv};
\addplot table [x=dali, y=fraction, col sep=comma, mark=*]{s_family.csv};
\addplot table [x=3d-blast, y=fraction, col sep=comma, mark=*]{s_family.csv};
\end{axis}
\end{tikzpicture}
      \caption{SCOPe\_fold}
      \label{figurelabel}
              \vspace*{6mm}
    \end{subfigure}%
    \quad
    \begin{subfigure}{0.5\textwidth}
     \centering
 \begin{tikzpicture}
 \centering
\begin{axis}[
ylabel near ticks,
xlabel near ticks,
xmin=0, 
xmax=3500, 
ymin=0, 
ymax=1,
xlabel={},
ylabel={}]
\addplot table [x=esp, y=fraction, col sep=comma, mark=*]{fold.csv};
\addplot table [x=fatcat, y=fraction, col sep=comma, mark=*]{fold.csv};
\addplot table [x=dali, y=fraction, col sep=comma, mark=*]{fold.csv};
\addplot table [x=3d-blast, y=fraction, col sep=comma, mark=*]{fold.csv};
\end{axis}
\end{tikzpicture}
      \caption{SCOPe\_family}
      \label{figurelabel}
                    \vspace*{6mm}
    \end{subfigure}
    \caption{ROC curves plotted over false positive rate (x) and fractional coverage over the database (y), for Esperite (blue), DALI (brown), FATCAT (red), and 3D-BLAST (black).}
   \end{figure*}
\subsubsection{SCOPe classification correlation} Based on the existing classifications in the SCOPe classification database, we assessed the performance in replicating protein structural neighbors within SCOPe. We evaluate the receiver-operating characteristic (ROC) curve and precision-recall curve
(PRC) analysis. The receiver-operating characteristic curve describes the amount of true positives identified against the amount of false-positives. The precision-recall curve, as the name suggests, draws a correlation between the precision against the overall coverage of the classification database.
We used the area under the ROC curve (AUC) and the area under the PRC (AUPRC) as measures
of agreement with the SCOPe classification.

\subsubsection{Computing time benchmarks.}
We compare how efficient Esperite is compared to other filter-and-refine methods Esperite is being benchmarked to in performing an exhaustive structural $N x N$ comparison of the 2,623 entries in SCOPe\_FAMILY.
\begin{table}[]
\centering
\caption{Exhaustive structural comparison running time (s) for Esperite and other benchmark filter-and-refine methods}
\label{my-label}
\begin{tabular}{@{}ccccc@{}}
\toprule
\multicolumn{1}{l}{} & FragBag & TM-Align & SGM & \textbf{Esperite} \\ \midrule
Processing           & 1324    & 3409     & 3386                      & \textbf{992}      \\
Running              & 1657    & 6876     & 113                       & \textbf{1312}     \\
Total                & 2981    & 10285    & 3499                      & \textbf{2304}     \\ \bottomrule
\end{tabular}
\end{table}
\subsection{Correlation with Multiple Alignment with Translations and Twists (MATT)}
Since Esperite only provides a list of potential structural neighbors, a full-body multiple structural alignment tool is needed to maximize the functionality of the web-based environment. We integrated the tool Multiple Alignment with Translations and Twists (MATT), engineered by Menke et al. (2008) [17] into the Esperite kernel, where we are able to conduct full-body structural alignment after narrowing the high-dimensional protein structure vector space down to a quite precise list of potential structural neighbors.\\\\
\indent Yet, it is crucial to make the distinction to need full structural alignment, and offer full structural alignment to the user. Hence, we conducted a correlation study between Esperite and MATT to decide whether or not the web-based tool need to run full structural alignment to get as accurate results as a computationally expensive structural alignment tool like MATT.\\\\
\indent MATT uses a unit length (block) that is the set of $\alpha$-carbon atoms in between five and nine amino acid residues in a protein. Given a block $B$, $b_h$ is the very first residue and $b_t$ is the very last
residue across the block. For a pair of blocks $BC$, $T_CB$ is the minimum RMSD transformation to trigger the second
structure to align its $\alpha$-carbons to the first, and $RMS_T$
is the RMSD of the two blocks under $T$. Then for $BC$,
\begin{equation}
Score(BC)=-logP(RMSD_T)
\end{equation}
Since both the $p-$value and the Cosine distance is bounded by [0,1], and Cosine distance is our accuracy benchmark classifier, we used a sample size of $N=250$ hits for trials with 5 different unique proteins to measure Cosine distance hits, executed MATT on the hits, and averaged the values.
\begin{figure}
\centering
 \begin{tikzpicture}[trim axis left, trim axis right]
 \centering
\begin{axis}[
ylabel near ticks,
xlabel near ticks,
xmin=0, 
xmax=1, 
ymin=0, 
ymax=1,
xlabel={Cosine Distance},
ylabel={$p-$value}]
\addplot table [scatter, x=cos, y=matt, col sep=comma, mark=*]{corr.csv};
\addplot {x};
\end{axis}
\end{tikzpicture}
      \caption{$p-$value generated by Multiple Alignment with Translations and Twists plotted with Cosine Distance for the same sample size by Esperite. $R^2=0.87$.}
      \label{figurelabel}
    \end{figure}
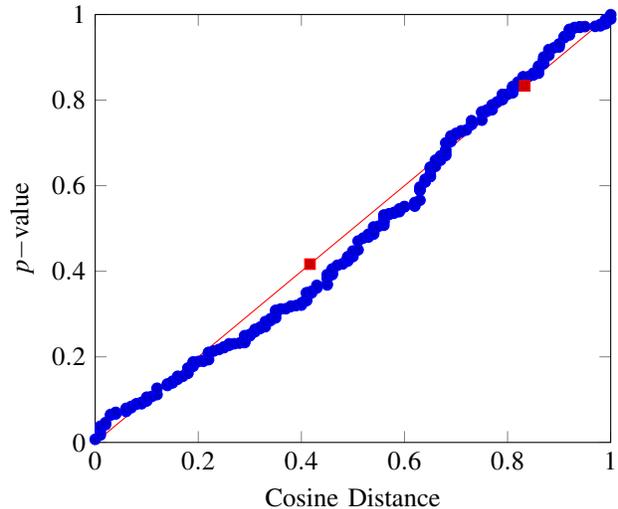

\section{ESPERITE: THE WEB SERVER}
We present Esperite, a web server devoted to applying our entropy-based clustering and search for protein structure neighbors in a user-centered,
web-based environment. Esperite serves as a real-time web-based tool through which structural neighbors can be identified in a fast, efficient and computationally expensive way. Using this algorithm, the bag-of-words representations of
the proteins in the Protein Data Bank are clustered for fast and efficient structure homolog
search in the database. Esperite is freely available to the public and is hosted by the MIT
servers. The web server
runs real-time commands for database search - so that the tool is available when the files
hosted on the server are damaged or deleted. Indeed, the files are retrieved directly from
the Protein Data Bank (PDB) and uploaded to the server without any prior modification.
Whenever the pdb file format and/or content changes in the database, the PDB file used
on the web server is updated and uploaded automatically. Every user starts a session with a
unique session ID which would then be used to reach results without entering parameters.
Once the session is over, the temporary files are deleted from the local host. All session IDs
are contained in a log file.
Esperite is up and running at http://esperite.csail.mit.edu and the source
code is also available on GitHub.
\begin{figure}[H]
      \centering
      \includegraphics[width=\linewidth]{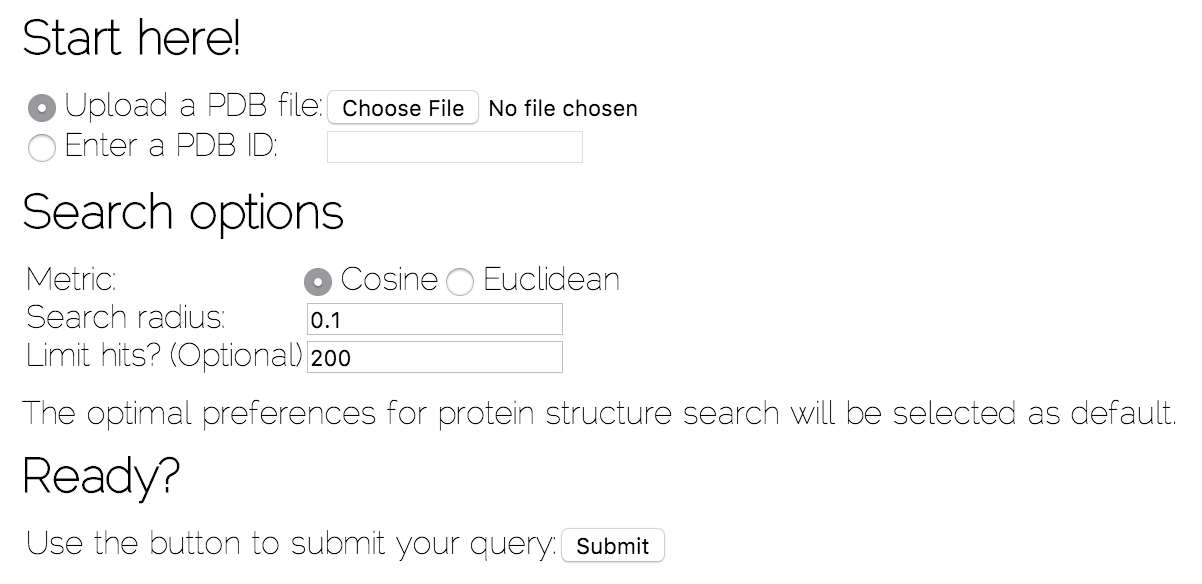}
      \caption{Query interface of our web-based ultrafast protein structure search tool, Esperite.}
      \label{figurelabel}
   \end{figure}
\subsection{Server Workflow.}
\subsubsection{Input} A PDB file contains the atomic coordinates for either a protein or other biomacromolecule.
Using different methods, structural biologists determine the location of each atom
of the molecule on the plane relative to one another, annotating and finally releasing the
resultant data as a PDB file to the public. The local shell command takes a PDB file as
an input for the structure search. The web server also allows for identification of the input
PDB file remotely from the Protein Data Bank itself: The user is presented with the option
of entering the PDB ID which is then be sent to the Protein Data Bank to locate the actual
PDB file. The PDB file is then uploaded temporarily to the local host and is deleted once
the user$’$s session ends. Once the file is uploaded to the local host, the database search shell
command is run from the server.\\
\begin{figure*}
      \centering
      \includegraphics[width=\linewidth]{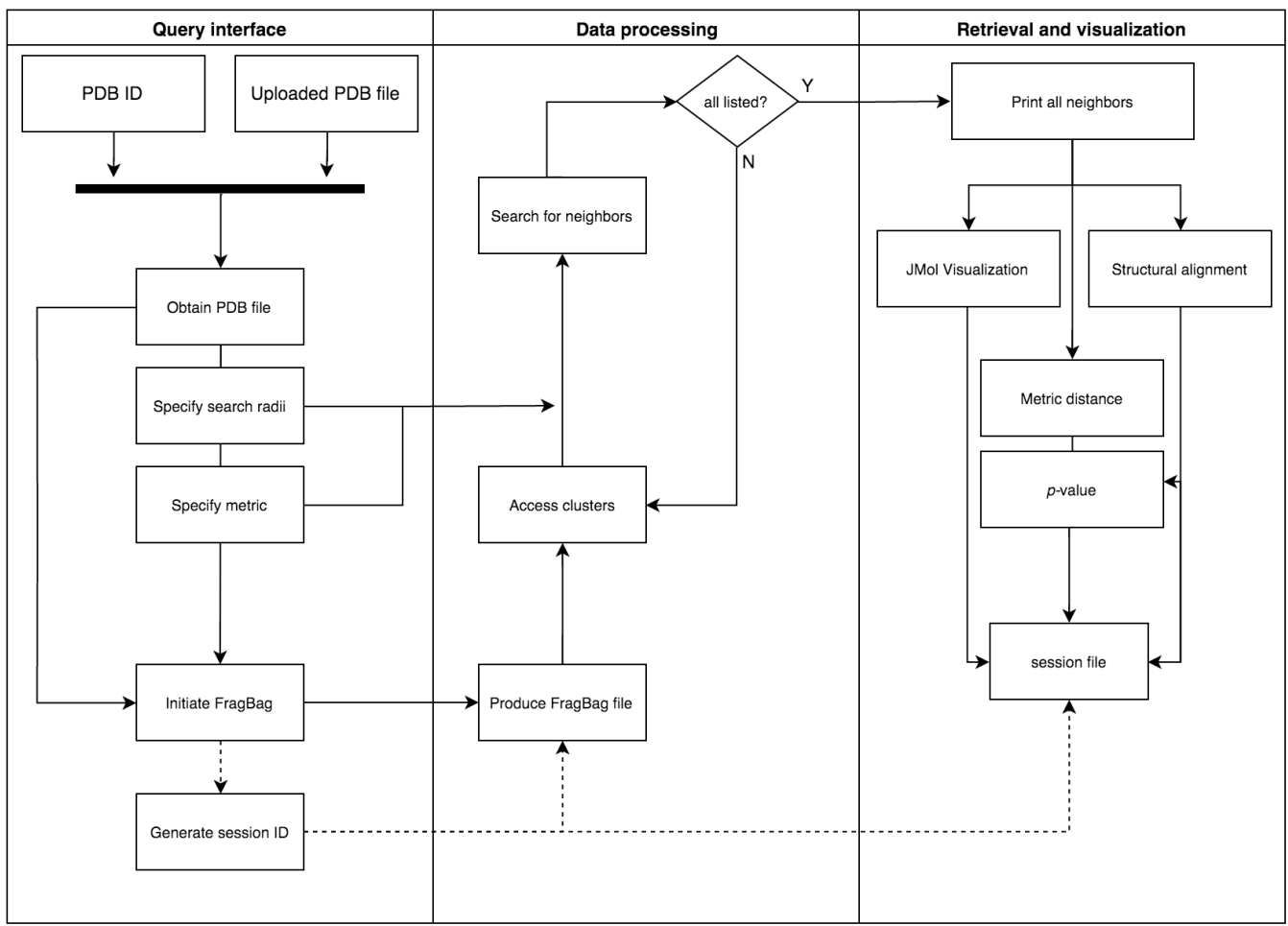}
      \caption{Query interface of our web-based ultrafast protein structure search tool, Esperite.}
      \label{figurelabel}
   \end{figure*}
\subsubsection{Search options} The web server offers options to customize the database search for structure
homologs of a PDB file. These options include: The metric that will be used to express
distance between two BOW vectors (Cosine distance or Euclidean distance), the initial search
radius $r$ (as an integer), the function that the initial search radius will be replaced with (current
options include $\sqrt{r}$, and $r/2$), and the depth of divisive clustering $d$ ($d$ also denotes
the level of the last cluster created). The user is also able to enter a function of their own
(provided that it complies with the syntax and the range of the dynamic radius $r'$). The user is also free to choose the default settings already set where $r = 10000$, $r' =\sqrt{r}$, used metric
is Euclidean distance and $d$ is not specified.\\
\subsubsection{Output}
After the job is submitted, the web server will display a self-refreshing page with
the task status until the job is completed. 
All files generated during the session are marked
with a unique ID for future reference, which is reported to the user.
The output window consists of a set of drop-down menus to display the PDB output file
of each hit, to run MATT for each hit and the input protein, and to see the PDB file in a
new tab. A dynamic AJAX JSMol window is also embedded to the results screen so the user
can access the real-time multiple flexible alignment quickly.
\section{CONCLUSION}
It has been recognized that predicting the function of a protein is key to understanding life at a molecular level. Sequence comparison, which is the most frequently used method to predict the function of a protein of known sequence but unknown function, has failed to identify relationships where the subject proteins may exhibit differences in function, regardless of the significant similarity in sequence. These relationships, hence, need to be investigated through another method which would ensure prediction of a function of an unannotated protein with
We introduced a novel entropy-based hierarchical framework to protein structure database clustering and search, allowing for linear-time similarity search even with an exponential increase in data. We rigorously proved that our approach scales linearly with the entropy of the database, thus sublinearly with the database itself. \\\\
\indent Given the drastic exponential increase in size of the Protein Data Bank, the biggest protein structure database online, the need for methods to scale in a sublinear proportion with the increase of data arises. We present an approach through which we are able to exploit the redundancy present in the Protein Data Bank, which has a plausible tendency to exhibit low-dimensionality on a local level.\\\\
\indent The main motivation behind the applicability of our approach is the minimal metric entropy and fractal dimension the protein structure database exhibits, which is demonstrated by the scaling behavior of the Cosine and Euclidean distances with increasing initial cluster radius.\\\\
\indent Furthermore, we contribute to the arising interest and knowledge on already existing filter-and-refine protein structure alignment methodology by a number of ways:
\begin{itemize}
\item To our knowledge, our approach is the \textbf{fastest} filter-and-refine protein structure alignment method to date, outperforming even state-of-the-art full-body structural alignments by a large margin.
\item To our knowledge, it is the \textbf{most accurate} filter-and-refine method that preserves the accuracy of the state-of-the-art full-body structural alignments to date.
\item It is the \textbf{only} approach to date whose time and space complexity bounds can be mathematically justified without any experimental background.
\item It is the \textbf{only} approach to scale sublinearly with the protein structure database to date, allowing for control over the increase of the database.
\item It is the \textbf{only} approach that allows for the discovery and investigation of structural neighbors of a newly-discovered protein, allowing for interesting biological elucidation.
\item It is the \textbf{only} approach that can construct structural neighbor families and superfamilies \textit{ab initio}, i.e., with no prior training with or starting from experimental data. 
\item The web server Esperite is the \textbf{only} protein structure alignment server who offers full flexible structural alignment using hits from a filter-and-refine method.\\
\end{itemize}
When we discuss the problem of finding the structural neighbors of a protein while the amount of information that needs to be processed increases, we are inclined to stay in the concept of proteomics. Yet, our approach is easily implementable to and versatile for any other "big data" problem our society is facing today. Any -omics data that are bounded in an exponentially increasing rate can be controlled and harnessed using entropy-based hierarchical clustering.
\section{FUTURE WORK}
\subsection{Thorough benchmark classifications}
Esperite is currently being thoroughly investigated by the Computer Science and Artificial Intelligence Laboratory (CSAIL) at the Massachusetts Institute of Technology (MIT) and the Department of Computer Science at Tufts University Graduate School of Engineering, both of which have endorsed Esperite to be the protein structure discovery tool of the respective computational biology labs in order to conduct more thorough benchmark classifications and detailed engineering of the tool.
\subsection{Machine learnable fold space}
\indent Corral-Corral et al. (2015) raise an important question: Given two representative data points of real-world objects, would the existence of a proximal similarity based on the dimensionality of the high-dimensional vector space in which the data points are clustered imply that the aforementioned two real-world objects are, in fact, similar? [43]
\begin{figure}
      \centering
      \includegraphics[width=0.7\linewidth]{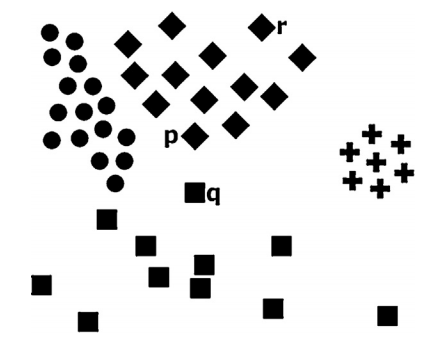}
      \caption{A visual depiction of the problem initially raised by Corral-Corral et al. (2015), where proximity in a high-dimensional representative vector space does not necessarily imply membership to a specific class. Rhombus class members $p$ and $r$ has a greater distance in between than that of $p$ and square class member $q$. This high-dimensional vector space is already clustered, and intuitively, we note that any characteristic of the given cluster is highly effective on the proximal errors a high-dimensional vector space can yield. Reprinted from [44].}
      \label{figurelabel}
   \end{figure}
 In order to begin to address this very frequently appearing issue in clustering and similarity search, Corral-Corral et al. (2015) posit that the lower and upper boundaries by which the proteins with different folds are surrounded can be investigated and numerically obtained via the training from empirical data and machine learning approaches.
\subsection{Working towards predicting the function of uncharacterized proteins}
We have established earlier that one of the biggest advantages of our entropy-based clustering and search method for protein structure comparison is that we are able to find structurally similar proteins to a protein of newly-discovered structure. Esperite is able to find structural neighbors \textit{ab initio}, so no empirical data on the structure of the query protein is needed. The Protein Data Bank is filled with proteins with newly-discovered structures every day, and Esperite appears to be the optimal utility to begin to identify and/or predict the function of various proteins.\\\\
\indent Using Esperite, we identified four uncharacterized proteins whose function can potentially be elucidated by further investigation. On the protein structural neighbor pairs shown in Figure 10, we continue our investigation at the New Mexico Highlands University (NMHU) in Las Vegas, NM, using nuclear magnetic resonance spectroscopy.
\begin{figure*}
\centering
\begin{subfigure}{0.40\textwidth}
\centering
   \includegraphics[width=\linewidth]{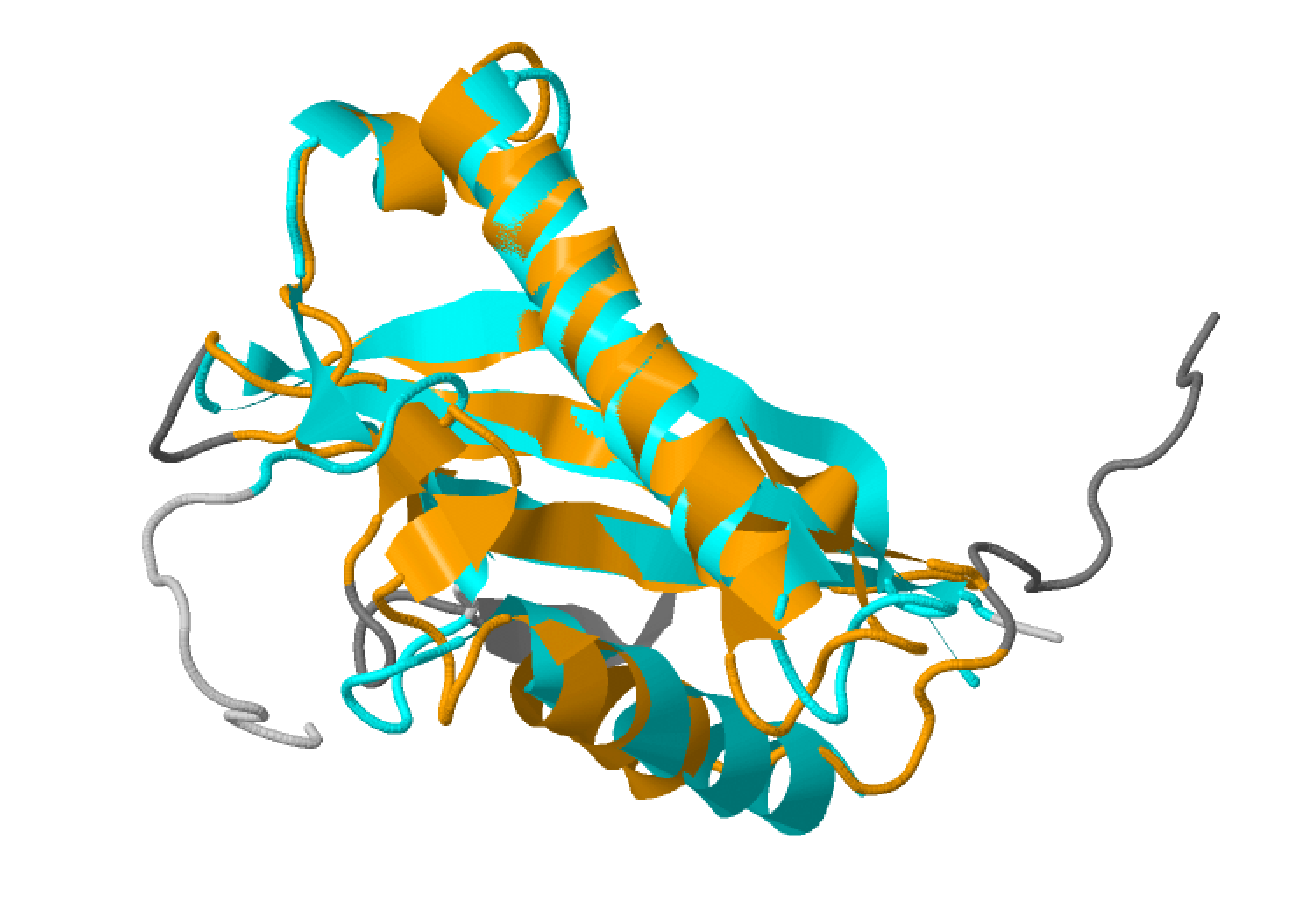}
      \caption{Uncharacterized protein (1IVZ) and SEA domain of transmembrane protease (2E7V)}
                    \vspace*{3.5mm}
    \end{subfigure}%
    \qquad
    \begin{subfigure}{0.40\textwidth}
        \centering
   \includegraphics[width=\linewidth]{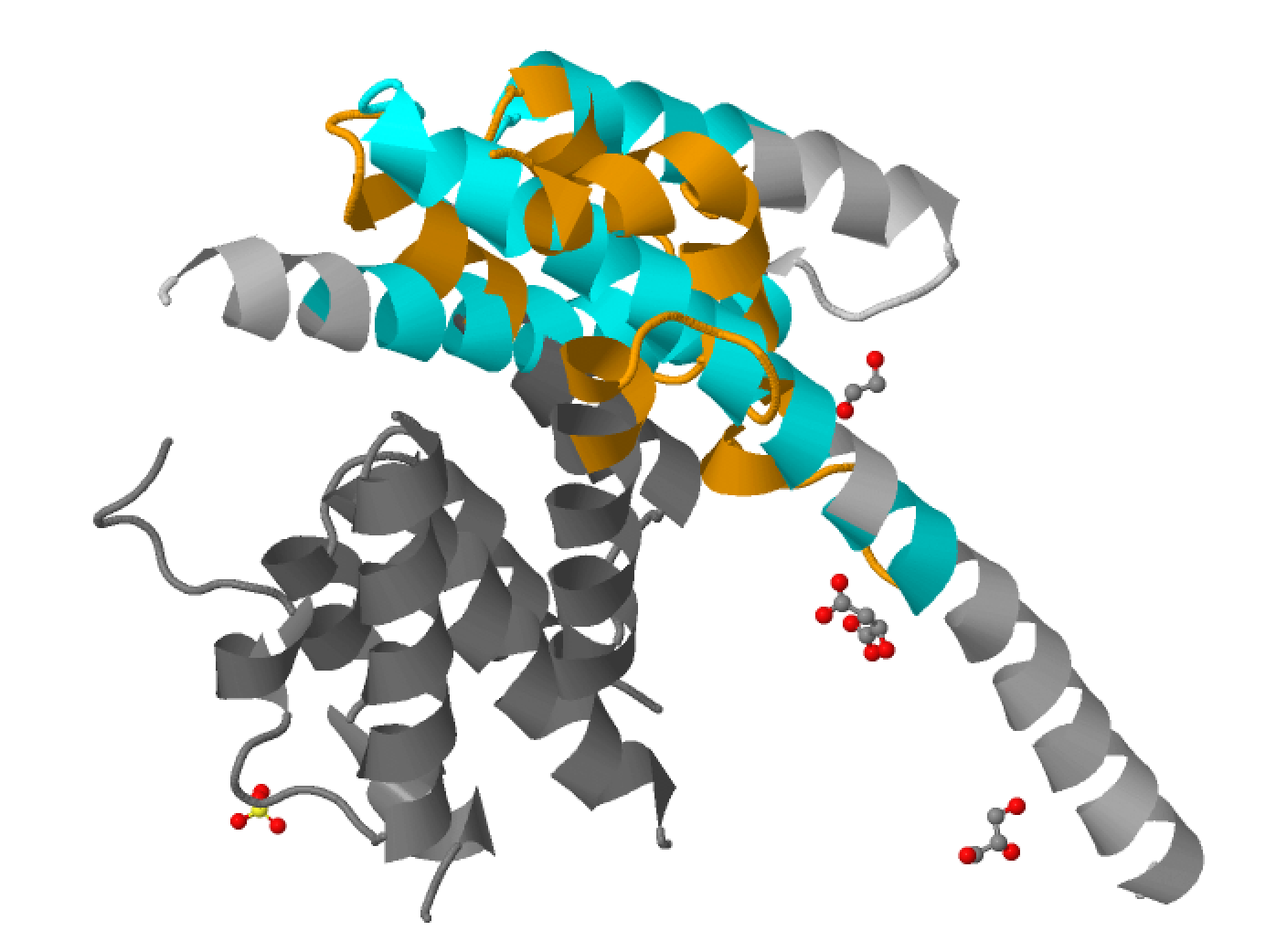}
      \caption{Uncharacterized protein (3B4Q) and chorismate mutase in complex with malate (3RMI)}
     \vspace*{6mm}
    \end{subfigure}
    \begin{subfigure}{0.40\textwidth}
        \centering
   \includegraphics[width=\linewidth]{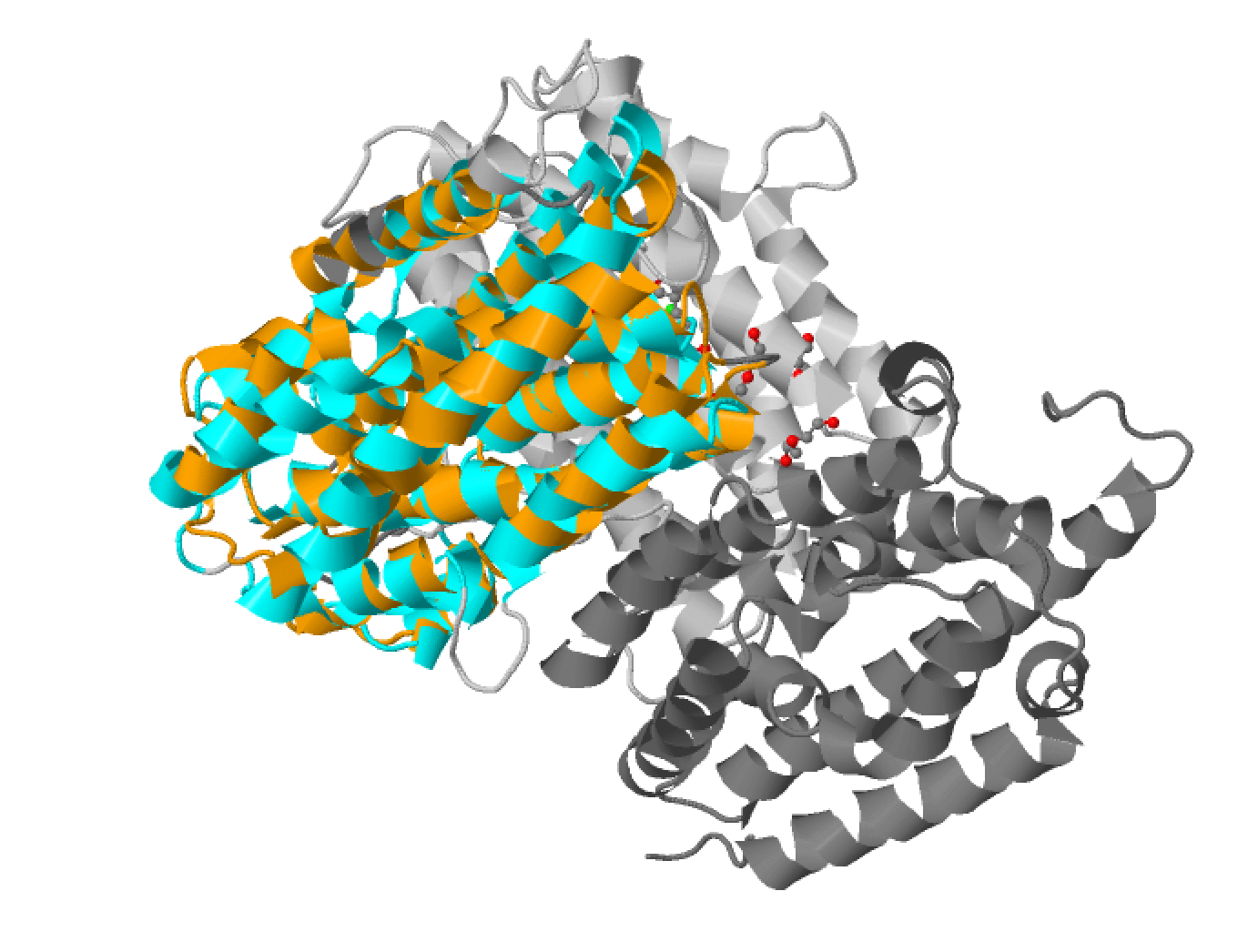}
      \caption{Uncharacterized protein (3DDE) and PqqC Active Site Mutant Y175F in Complex with PQQ (3HLX)}
    \end{subfigure}%
        \qquad
    \begin{subfigure}{0.40\textwidth}
    \centering
   \includegraphics[width=0.7\linewidth]{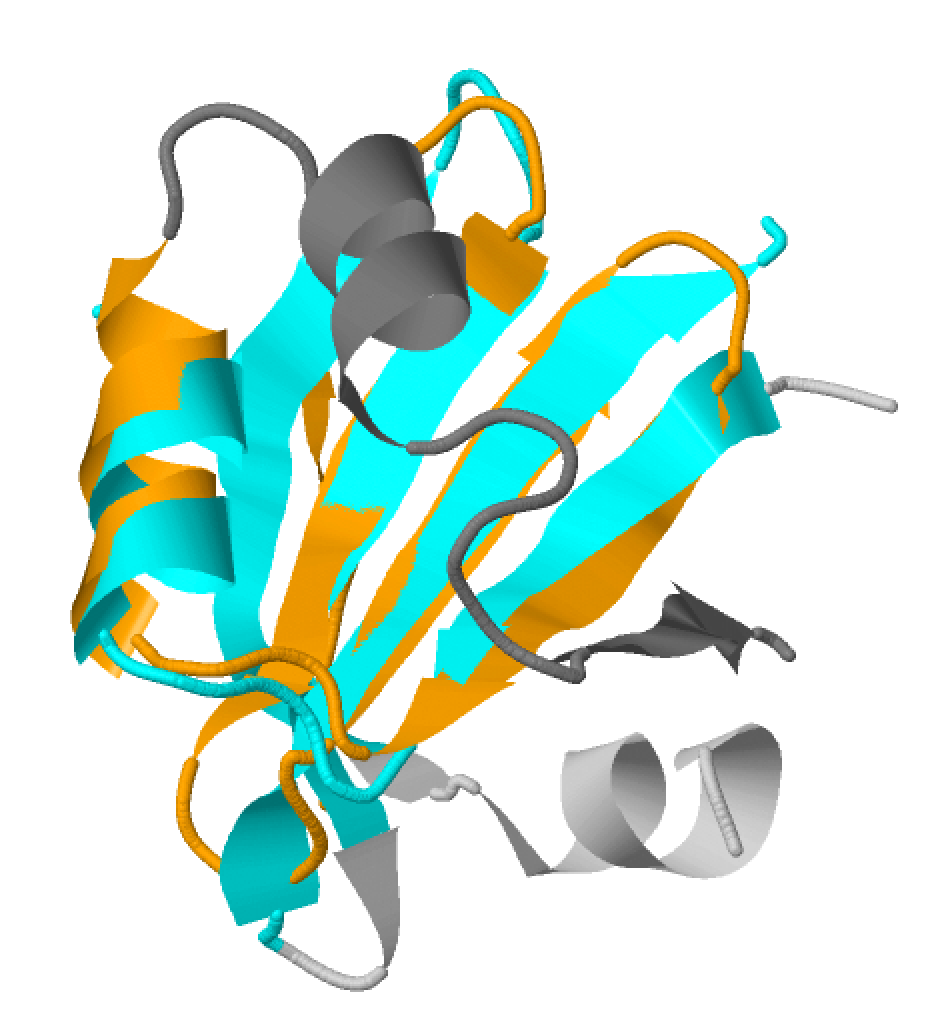}
      \caption{Uncharacterized protein (4RGI) and m7GpppX diphosphatase (5BV3)}
    \end{subfigure}
    \caption{Four uncharacterized proteins (1IVZ, 3B4Q, 3DDE, and 4RGI) and their closest structural neighbors (2E7V, 3RMI, 3HLX, and 5BV3, respectively), printed as output by Esperite (Cosine distance between any one of the pairs were 0.0), and to be investigated further through protein nuclear magnetic resonance spectroscopy.}
\end{figure*}
\section{AVAILABILITY}
All of our software is available in GitHub, under the GNU General
Public License, and our tools can be integrated into existing frameworks and the code and methodology can be incorporated into other software.\\\\
\indent Esperite is up and running at http://esperite.csail.mit.edu.
\section*{APPENDIX A: THEORETICAL FOUNDATIONS}
\subsection*{Evaluation of time complexity.}
To facilitate and streamline the analysis of time complexity, we consider the high-dimensional vector space as a set $S$ of the collection of $n$ unique points:
\begin{equation}
D_{S}(q, r) = {p \in S : ||q - p|| < r}
\end{equation}
The problem at hand then becomes to compute $D_{S}(q, r)$ for a query protein $q$ and a given non-zero radius $r$. It is important to note that a distance metric, particularly, is needed to only avoid a non-zero loss in sensitivity.\\\\
\indent
Initially, a random cluster center $k$ is defined in this high-dimensional vector-space $S$. The points around this cluster center $k$ within a user-defined radius $r_{c}$ are then clustered and assigned to the cluster center $k$. Iteratively, a set $C$ of $k$ cluster centers are defined so that there are no clusters with a cluster radius greater than a user-defined radius $r_{c}$. We assign and note each cluster to its center, so the set $C$ is also the exhaustive set of clusters in the high-dimensional vector space $S$. For a given task of similarity measurement between a query protein $q$ and all data points within distance $d$, the approach consists of $n$ iterative steps, combined with two real-time variables $r_{c}$ and $r$ that change values according to another user-specified parameter $p_{r}$, where
\begin{equation}
r=p_{r}r_{c}
\end{equation}
For ease of analysis, we will work on a hierarchical framework where $n=2$, i.e., only two layers of clustering is done in the high-dimensional vector space $S$ (note that a deeper clustering will result in a depth of $kn$, where $k$ is a non-zero integer, thus scaling linearly with the running time of the search). The overall search is then split into 2 stages of searches, with radii of $r+r_{c}$ and $r$, respectively. The similarity task of the first stage of the search, or the \textit{coarse search}, is then $D_{C}(q, r + r_{c})$, and the union of all defined clusters $C_{k}$ with center $k$ for the radius $r+r_{c}$ is then
\begin{equation}
U_{C}=\bigcup _{C_{k} \in D_{C}(q, r + r_{c})}C_{k}
\end{equation}
Triangle Inequality, which all distance metrics obey, implies that the second stage of the search, or the \textit{fine search}, $D_{F}(q, r) \subseteq U_{C}$. Intuitively, as $U_{C} \subseteq S$,
\begin{equation}
D_{F}(q, r)=D_{C}(q, r)
\end{equation}
So the same defined clusters $C_{k}$ with centers $k$ and radius $r+r_{c}$ in the coarse search can be used in fine search with radius $r$. \\\\
\indent
In an entropy-based construction of the dataset, the distance function used is only a metric to avoid disobeying the Triangle Inequality. It has been noted by Yu et al. (2015) that the use of many distance functions are plausible, yet decrease the amount of sensitivity the approach exhibits. More particularly, if a triplet of data points $T_{\alpha}$ over all triplets $T_{S}$ in the high-dimensional vector space $S$ do not satisfy the Inequality, the sensitivity is then $1 - \frac{T_{\alpha}}{T_{S}}$ [28]. As experimentally shown in the results, this loss is infinitesimal, and inclined to lose significance as $r+r_{c} \rightarrow \infty$.\\\\
\indent
If the fractal dimension of the high-dimensional vector space $S$ is low, the $n$-stage hierarchical clustering approach has a running time increase that is linearly proportional with the metric entropy of the database. We note that this database, along with any newly-added points to the database, can also be stored in a space complexity linearly proportional to the metric entropy of the database.
\subsection*{Bounds of complexity.}
\textbf{Definition 1.} Let $S$ be a high-dimensional vector space, $D$ a subset of $S$, and $r$ a non-zero radius. The \textit{metric entropy} $E_{r}(S)$ of the high-dimensional vector space $S$ is then the minimum number of points $p_{1}, p_{2}, ..., p_{n}$ so that spherical clusters $S(p_{1}, r), S(p_{2}, r) ..., S(p_{n}, r)$ include all points in $S$ [44]. \\\\
\indent
\textbf{Definition 2.} Given any high-dimensional vector space $S$, the Hausdorff-Besicovitch dimension is
\begin{equation}
dim_{H}=\lim_{r\to 0}\frac{\log E_{r}(S)}{\log 1/r}
\end{equation}
Intuitively, we note that for all cases, $dim_{H} = 0$ since $S$ is finite and countable. We then utilize a more narrow definition of fractal dimension located around a distance scale with large radii:\\\\
\indent
\textbf{Definition 3.} Given any high-dimensional vector space $S$, and a radius scale $[r_{1},r_{2}]$, the \textit{fractal dimension} is
\begin{equation}
dim_{f}=\lim_{r\to r_{2}}\frac{\log \frac{E_{r}(S)}{E_{r_{1}}(S)}}{\log \frac{r_{1}}{r}}
\end{equation}
Intuitively, we note that after the initial clustering with radii $r_{c}$ and the construction of a set $C$ of $k$ cluster centers in the high-dimensional vector space $S$, $k \geq E_{r_{c}}(S)$, since the criterion that no cluster has a radius greater than $r_{c}$ is satisfied. Thus, set of cluster centers $k$ is bounded by metric entropy $E_{r_{c}}(S)$ of the high-dimensional vector space $S$.\\\\
\indent Still following the specific case of $n$-stage hierarchical clustering approach where $n=2$, we note that:
\begin{itemize}
\item For a given query $q$, the coarse search will be exhaustively completed only after $k$ comparisons.
\item The fine search is bounded in the union $U_{C}$ in (5), which is the union of clusters $r+r_{c}$ away from query $q$. Thus, the running time for the $n=2$ search overall is $\mathcal{O}(k+|U_{C}|)$, as the sum of running times for both the coarse and the fine search.
\item $U_{C} \subset D_{C}(q, r + 2r_{c})$ by the triangle inequality (proof is trivial); thus, $U_{C}$ is bounded by $D_{C}(q, r + 2r_{c})$.
\item Recall Definition 3. Given the non-rigorous definition of fractal dimension, we note that for a scale of radii $[r_{\alpha}, r_{\beta}]$, and an increase from $r_{\alpha}$ to $r_{\beta}$, any newly-discovered point will be covered by fractal dimension $d$ in a running time of $\mathcal{O}(\frac{r_{\beta}}{r_{\alpha}})^d$.
\end{itemize}
Thus, the overall time complexity for $n=2$ is
\begin{equation}
T(n)=\mathcal{O}(k+D_{C}(q, r)(\frac{r + 2r_{c}}{r})^d)
\end{equation}
Note that for asymptotically small values of fractal dimension and a linear output size, the running time converges to 
\begin{equation}
T(n)=\mathcal{O}(n\,log\,n)
\end{equation}
as $k$ is linearly scaled with entropy, thus logarithmically with the dataset; and the output size $n$ is linear.
\section*{APPENDIX B: MOTIVATION FOR DIVISIVE HIERARCHICAL CLUSTERING}
Two distinct methods can be used for hierarchical clustering: agglomerative, and divisive.
Agglomerative clustering is when each data point is initially accepted as a cluster within itself, and these clusters are merged using the similarity quantification via distance metrics.
Clustering continues until the high-dimensional vector space $S = 1$.\\\\ 
\indent Divisive clustering can
be thought of as the inverse: The high-dimensional vector space $S$ is initialized as the first cluster,
and closer data points in the cluster create a new sub-cluster. Divisive hierarchical clustering continues until every single data point in $S$ is a cluster within itself.\\\\
\indent Because the algorithm starts with the high dimensional  vector space $S$ being the
initial cluster, and ends when all data points are clusters within themselves or user-defined $d$
equals timer $t$, the system is based on flat divisive hierarchical clustering. Although these two
methods may seem like they are supposed to replicate results when timer $t$ is reversed, when
the distance metric and the similarity criterion are introduced, using divisive hierarchical
clustering is proven to be more optimal:
\begin{itemize}
\item Assume you have a cluster center $k$ and a radius $r$ when $t = 0$ and $R(r) = 2r$. Every data
point $n$ where $D_{nk} = r-s$ where $s \rightarrow 0$ is included in cluster $C_{k}$.
\item  If this cluster
center $k$ is $2r-s$ further away from another cluster center $l$, then when $t = 1$, $C_{k}$ and $C_{l}$
should be merged. 
\item At this point, we want the data points which are members of $C_{l}$ to be
less than or equal to $2r$.
\end{itemize}
 Yet when we merge the two clusters, $n$ is almost $3r$ away from $l$ but
is still a member of $C_{l}$.\\\\
\indent Agglomerative clustering is flawed in similarity search; it is not error-free for radius $r$, thus the cluster centers are not entirely reliable as they include more distant points in each step.
The advantage of starting clustering from a high-dimensional vector space $S$ is that since each level of clustering
happens only within the clusters from the previous level, clusters that have elements
that are too far from one another are not merged. It is also a more succinct implementation
on top of the existing initial clustering.
\section*{ACKNOWLEDGMENTS}
The author thanks Dr. Bonnie Berger and Dr. Noah Daniels at Computer Science and Artificial Intelligence Laboratory (CSAIL) at the Massachusetts Institute of
Technology (MIT) for their mentorship
and guidance, Dr. Semsettin Turkoz, Mr. and Mrs. Mustafa Basar Arioglu, and Mr. Kemal
Tunc for their sponsorship, Joseph Dexter and Marie Herring for their suggestions and
support, Andrew Jin and Daniel Yang for their feedback, Andrew Gallant and Yun William
Yu for FragBag implementation, and the related groups for making the following packages
available: the Jmol/JSMol package (http://jmol.sourceforge.net), Go Programming
Language (http://golang.org), and the PDB (http://rcsb.org/pdb).


\end{document}